%% file: paper_sfre_dpp.tex
\documentclass[%
 reprint,
nofootinbib,
 amsmath,amssymb,
 aip,
floatfix,
longbibliography,
]{revtex4-2}

\usepackage{graphicx}% Include figure files
\usepackage{subcaption}
\usepackage{dcolumn}% Align table columns on decimal point
\usepackage{bm}% bold math
\usepackage[english]{babel}
\usepackage{upgreek}
\usepackage{ifthen}
\usepackage{subcaption}
\usepackage{float}
\usepackage{comment}
\usepackage{tikz}
\usetikzlibrary{shapes.geometric}
\usepackage{pgfplots}
\pgfplotsset{compat=1.17}

\usepackage{diagbox}

\input{macros.tex}

\newcommand{\e}[1]{\mr{e}^{#1}}

\newcommand{\vrel}{v_\mr{rel}}
\newcommand{\vtot}{v_\mr{tot}}
\newcommand{\prel}{p_\mr{rel}}
\newcommand{\ptot}{p_\mr{tot}}

\newcommand{\fm}{f_\mathrm{M}}

\newcommand{\nuii}{\nu_\mr{ii}}

\renewcommand{\sone}{\alpha}
\renewcommand{\stwo}{\beta}

\newcommand{\vthone}{v_{\sone,\mathrm{th}}}
\newcommand{\vthtwo}{v_{\stwo,\mathrm{th}}}

\newcommand{\lthsone}{\lambda_{\sone,\mathrm{th}}}

\newcommand{\subfiglabelsizept}{12}
\newcommand{\subfiglabelleadingpt}{12}

\newcommand{\subfigpanelxshift}{-0.7cm}

\usepackage{mathptmx}
\usepackage{etoolbox}
\usepackage{ragged2e}

\usepackage[font=small]{caption}

\makeatletter
\def\@email#1#2{%
 \endgroup
 \patchcmd{\titleblock@produce}
  {\frontmatter@RRAPformat}
  {\frontmatter@RRAPformat{\produce@RRAP{*#1\href{mailto:#2}{#2}}}\frontmatter@RRAPformat}
  {}{}
}%
\makeatother
\begin{document}

%\preprint{APS/123-QED}
\preprint{AIP/123-QED}

\title{Asymptotic behavior of the shear flow reactivity enhancement effect}

\author{Henry Fetsch}
 \email{hfetsch@princeton.edu}
\author{Nathaniel J. Fisch}%
\affiliation{Department of Astrophysical Sciences, Princeton University, Princeton, NJ}

\date{\today}

\begin{abstract}

Fusion reactivity is enhanced in the vicinity of strongly sheared flow due to the tendency of fast ions near the Gamow peak to travel long distances between collisions, thereby sometimes crossing gradients in the background flow and attaining a velocity boost relative to the thermal background. This ``shear flow reactivity enhancement effect'' (SFRE) allows turbulent kinetic energy on fine spatial scales to contribute to fusion reactivity before thermalizing, which, remarkably, enables ignition of some inertial confinement fusion (ICF) hot spots under conditions where fully thermalized plasma would fail to ignite. 
The size of the SFRE is a consequence of the dramatic scale separations distinguishing thermal ions, which govern fluid quantities, and fast ions, which govern fusion reactivity. It is demonstrated in this work that, as the Gamow energy increases relative to the thermal energy, the SFRE in unmagnetized plasma becomes asymptotically large compared to hydrodynamic effects such as viscous dissipation. An asymptotic formula is derived in this limit, quantifying the SFRE for reactants of disparate masses and charge states.

\end{abstract}

%PoP impact statement: (Accompanying paper to an APS-DPP invited talk.) Fusion reactivity is typically treated as a function of the ion temperature only. However, a recently identified "shear flow reactivity enhancement effect" (SFRE) demonstrates that flows in a fusion plasma can contribute directly to reactivity due to the propensity of fast ions (far more so than their thermal counterparts) to travel long distances across flow gradients and thereby develop an enhanced tail. Previous works (which provided the focus of the accompanying invited talk) introduced the effect, derived an approximate analytical theory for the reactivity enhancement factor in turbulent plasma, and validated the effect numerically. However, the crucial asymptotic scalings of the SFRE have not previously been presented in any detail. These fundamental aspects of the SFRE are the subject of this work. In addition, the reactivity enhancement is derived for pairs of reactants with different masses and charge states, which is of particular importance in astrophysical fusion plasmas. 

\maketitle

\section{Introduction}
\label{sec_intro}

%In fusion plasmas, ions are repelled by their mutual positive charges so that, in the classical limit, ions with energies below the MeV range never approach each other closely enough to undergo fusion. Quantum tunneling allows some ions colliding with lower relative energies, typically in the tens of keV range, occasionally to pass through the Coulomb barrier and fuse. 
%Because the probability of tunneling is small for ions with energies well below the energy of the Coulomb barrier, the rate of fusion reactions remains low relative to other collisional processes. In thermonuclear plasma, increasing the ion temperature leads to more energetic collisions and can dramatically increase the fusion rate. 

To produce large fusion yields, plasmas must be heated to high temperatures because only pairs of energetic ions colliding with large relative velocities have a significant chance of tunneling through the Coulomb barrier and fusing. 
Under conditions characteristic of most laboratory fusion devices, and most astrophysical systems, the probability of fusion between thermal ions is small. It is instead a small population of fast ions that drives most fusion reactions \cite{Fetsch_Fisch_2025a,Bosch_Hale_1992}. 
As a result, fusion reactivity is sensitive to the tail of the ion distribution, meaning that even relatively small deviations from a Maxwellian can produce large changes in reactivity. Several non-thermal distributions are known to offer significantly higher reactivity than Maxwellians of the same energy, including anisotropic ``multiple-temperature" Maxwellians\cite{Kolmes_Mlodik_Fisch_2021,Xie_Tan_Luo_Li_Liu_2023}, beam-target distributions\cite{Higginson_Link_Schmidt_2019}, and distributions with enhanced tails \cite{Garbett_2013,Fetsch_Fisch_2025a,Fetsch_Fisch_2025b,Ye_Zhang_Wan_2025}. 
In practice, however, collisions and instabilities act quickly to relax non-Maxwellian distributions toward local thermal equilibrium; the small size of the fusion cross section relative to the cross section for Coulomb collisions under most conditions means that relaxation occurs much more quickly than fusion \cite{Rider_1997}. 
In most high-yield fusion devices, therefore, the dominant fraction of the yield is generated in plasma with nearly Maxwellian ion velocity distributions.

This rapid relaxation to local equilibrium justifies the widely used approximation in which fusion reactivity is taken to be a function only of local thermodynamic variables, such as temperature and density. Allowance is sometimes made for distinct ion and electron temperatures -- the so-called \textit{hot-ion mode} with ion temperatures greater than electron temperatures is highly desirable for achieving economical nuclear fusion \cite{Clarke_1980} -- but energy residing in other degrees of freedom, such as flows and radiation, is typically taken to have no direct effect on reactivity. 
Under some circumstances, however, this simple picture breaks down, producing remarkable non-equilibrium effects. 
For example, it is well known that waves resonating with fast ions can generate an enhanced tail, meaning that ion cyclotron resonance heating (ICRH) can boost fusion rates \cite{Harvey_McCoy_Kerbel_Chiu_1986,Frank_Wright_Rodriguez-Fernandez_Howard_Bonoli_2024} in magnetic confinement fusion (MCF). 
Waves can furthermore be used to divert the energy of alpha particles to ions rather than to electrons, which would ordinarily receive most of the alpha heating \cite{Fisch_Rax_1992}. Known as \textit{alpha channeling}, the maintenance of this non-equilibrium state is advantageous in a variety of MCF devices \cite{Fisch_Herrmann_1994,Fisch_Herrmann_1999}. 
Hydrodynamic shocks\cite{Mannion_et._2023,Jeet_et_2023}, as well as some kinetic instabilities \cite{Liu_Li_Yao_Lei_Zhou_Zhu_He_Qiao_2021}, can drive ion distributions out of equilibrium and produce a higher fusion reactivity in inertial confinement fusion (ICF) devices. 
In hot plasma near a cold boundary, reactivity is reduced as fast ions are preferentially lost from the burning region, which can lead to anomalously low yields in ICF\cite{Petschek_Henderson_1979,Molvig_Hoffman_Albright_Nelson_Webster_2012,Albright_Molvig_Huang_Simakov_Dodd_Hoffman_Kagan_Schmit_2013,Davidovits_Fisch_2014,McDevitt_Tang_Guo_2017}. 

It was recently shown\cite{Fetsch_Fisch_2025a} that, counterintuitively, even in the absence of waves or interfaces, hydrodynamic motion alone is sufficient to enhance fusion reactivity well above its thermal value in some cases. This ``shear flow reactivity enhancement'' effect (SFRE) stems from the long mean free paths of the fast ions responsible for fusion reactions, which make these ions sensitive to flow gradients that are negligible on the scales relevant to thermal particles \cite{Fetsch_Fisch_2025a}. As a result, in plasmas containing small-scale turbulent flows, fusion reactivity ceases to be a function of ion temperature alone and becomes, instead, a functional of the turbulent energy spectrum \cite{Fetsch_Fisch_2025b}. 
Remarkably, the SFRE can allow ignition under some conditions where fully thermalized plasma would not ignite. This could have practical benefits in the design of ICF experiments, where flows might be driven on short length scales at late stages of an implosion to boost reactivity immediately prior to ignition \cite{Fetsch_Fisch_2026}. 
A further advantage of partitioning some implosion energy into fine-scale turbulence, rather than into thermal energy, is that a lower temperature during compression can reduce losses due to radiation and thermal conduction\cite{Davidovits_Fisch_2016a,Davidovits_Fisch_2016b,Davidovits_Fisch_2019a}; substantial turbulent kinetic energy has been deduced at stagnation in Z-pinch plasmas, coinciding with less radiation during compression \cite{Kroupp_et_2018}.
Implications of the SFRE for ICF designs are discussed in Refs. \citenum{Fetsch_Fisch_2025a,Fetsch_Fisch_2025b,Fetsch_Fisch_2026}, including the order-of-magnitude advantages that might be attained by substituting thermal for turbulent energy in fast-ignition or shock-ignition designs. This work, by contrast, focuses on the fundamental nature of the SFRE as a consequence of the dramatic separation in plasma between the scales relevant to thermal ions and those relevant to ions that govern fusion reactivity.

%Achieving a substantial non-thermal advantage requires introducing a mechanism to maintain ion distributions out of equilibrium on time scales longer than the ion-ion collision time. 
%This can be accomplished, for example, by driving waves that preferentially accelerate ions with velocities greater than the thermal velocity, generating an enhanced tail. 
%A recently identified ``shear flow reactivity enhancement effect'' (SFRE) offers a novel means of sustaining non-thermal tail distributions by small-scale hydrodynamic turbulence, rather than by waves.

The remainder of this work is organized as follows. Section~\ref{sec_background} reviews the physics of fusion reactivity and fast-ion collisionality in plasmas and demonstrates that preferentially enhancing the tail of the ion distribution function increases reactivity more efficiently than simply heating the ions. Next, \S\ref{sec_sfre} illustrates the core physics of the SFRE and shows through simple asymptotic arguments that sheared flows can multiply reactivity by several times, even at modest flow velocities. 
A quantitative version of this argument is given in \S\ref{sec_turb_reac}, where an asymptotic formula is derived for the reactivity of turbulent multi-ion plasmas. In \S\ref{sec_visc_diss}, viscous effects are compared to the SFRE and shown to become asymptotically small when the Gamow energy is large relative to the thermal energy. Finally, \S\ref{sec_disc} discusses implications of the SFRE for the design of ICF experiments.  

\section{Background}
\label{sec_background}

\subsection{Fusion reactivity}

A fusion reaction is a binary process whose cross section $\sigma$ depends on the relative speed $\vrel$ of the colliding ions. 
By convention, $\sigma$ can be written as
\begin{equation}
  \label{eq_sigma_S}
  \sigma(\vrel) = S(\vrel) \frac{A \e{-\sqrt{2 E_G/\mu \vrel^2}}}{\vrel^2}  ,
\end{equation}
where $A$ is a constant, $\mu$ is the reduced mass of the reactants, and $S$ is a dimensionless factor that depends on $\vrel$. The exponential term in \eqref{eq_sigma_S} describes the barrier-penetration probability in terms of the Gamow energy $E_\mr{G}$, defined as
\begin{equation}
  \label{eq_E_G_defn}
  E_\mr{G} = 2\pi^2 \frac{\mu Z_1^2Z_2^2 e^4}{\hbar^2}
\end{equation}
where $Z_1$ and $Z_2$ are the charge states of the reactants, $e$ is the elementary charge, and $\hbar$ is the reduced Planck constant.
In \eqref{eq_sigma_S}, the $1/v^2$ factor is a kinematic term originating, heuristically, from the fact that ions with shorter de Broglie wavelengths are less likely to collide head-on. The factor $S(\vrel)$ absorbs all remaining velocity dependence due to the particular nuclear physics of the reaction of interest. Away from resonances, $S$ is a slowly varying function and can sometimes be approximated as a constant to isolate the dominant exponential behavior of the cross section.

The fusion reactivity can be written as ${\avg{\sigma v} = \Sigma[f_\sone,f_\stwo]/(1 + \delta_{\sone,\stwo})}$, where $\delta_\mr{\sone,\stwo}$ is the Kronecker delta between species $\sone$ and $\stwo$, $\Sigma$ is a bilinear functional defined as
\begin{equation}
  \label{eq_reac_def}
  \Sigma[f_\sone,f_\stwo] = \iint d^3 v d^3 v' \sigma(\vrel)\vrel f_\sone(\bs v) f_\stwo(\bs v'),
\end{equation}
and $f_\sone$ and $f_\stwo$ are distribution functions.

The reactivity can be straightforwardly estimated when the reactants are Maxwellian, \textit{viz.} $f(v) = \fm(v)$, where 
\begin{equation}
  \label{eq_f_M_def}
  \fm(\bs v) = \frac{1}{(2\pi)^{3/2}\vth^3} \e{-\half |\bs v - \bs u|^2/\vth^2} ,
\end{equation}
$\bs u$ is the drift velocity (which is the same for both reactant species), and ${\vth = \sqrt{T/m}}$ is the thermal velocity for ions of temperature $T$ and mass $m$. Let $\bs p = (\bs v - \bs u)/\vth$ be the normalized peculiar velocity. Then, taking the case of a single reactant species for simplicity, \eqref{eq_reac_def} becomes
\begin{equation}
  \label{eq_reac_maxwellian}
  \avg{\sigma v} = \frac{A }{(2\pi)^3 \sqrt{2}\vth} \iint d^3 \ptot d^3 \prel \frac{S(\vrel)}{\prel} \e{-b/\prel - \half \prel^2 - \half \ptot^2},
\end{equation}
where $b = \sqrt{E_G/\mu \vth^2}$; in this work, we will refer to $b$ as the ``Gamow parameter''. The velocities $\bs p$ and $\bs p'$ of the reacting particles have been transformed to the variables ${\bs \prel = (\bs p - \bs p')/\sqrt{2}}$ and ${\bs \ptot = (\bs p + \bs p')/\sqrt{2}}$ (so ${\vrel = \sqrt{2}\vth \prel}$). Note that, in \eqref{eq_reac_maxwellian}, terms involving $\ptot$ are independent of the fusion cross section; they are simply a Gaussian integral, which can be factored out and integrated exactly. In other words, when a pair of ions with velocities $\bs v$ and $\bs v'$ collide, their center-of-mass velocities form a Gaussian distribution, with a typical speed of $\vtot = \sqrt{2}\vth$.

\begin{figure}[t]
  \centering 
  \includegraphics[width=0.95\columnwidth]{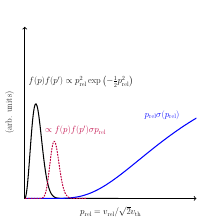}
  \caption{\justifying Distribution of relative velocities (black) for Maxwellian reactant distribution compared with a representative fusion cross section (blue). The reactivity integrand (purple) is dominated by a narrow window around the Gamow peak.}
  \label{fig_gamow_window}
\end{figure}

Because $E_G$ is large, typically lying in the MeV range or higher, $b$ is a large parameter. For example, for the deuterium-tritium (DT) reaction at $T = 3~\mr{keV}$, $b \approx 30$, whereas for the carbon-12-carbon-12 (\textsuperscript{12}C\textsuperscript{12}C) reaction at $T = 50~\mr{keV}$, $b\approx 550$. (Temperature is expressed in units of energy in this work.)
As a result, \eqref{eq_reac_maxwellian} is well suited to asymptotic evaluation. The exponential part of the integrand is rapidly varying compared to the $S(\prel)\prel$ prefactor and can be approximated as a Gaussian peaked at $\prel = \sqrt{2}p_* = b^{1/3}$ to find
\begin{equation} 
  \label{eq_reac_maxwellian_approx}
  \avg{\sigma v} \sim \sqrt{\frac{2}{3}}\frac{A b^{1/3}}{\vth} S\paren{ b^{1/3}\sqrt{2}\vth} \e{-\frac{3}{2}b^{2/3}} .
\end{equation}
The dependence of the reactivity integrand on a narrow region of velocity, the ``Gamow peak'' can be seen in Fig.~\ref{fig_gamow_window}. For Maxwellian reactants, the relative velocities form a Gaussian distribution peaked in the region around $\vrel = 2\vth$. In this region, the fusion cross section is small; pairs of thermal ions are unlikely to fuse with each other. For the parameters $b=50$ and $S=1$ used for illustration in Fig.~\ref{fig_gamow_window}, $\sigma$ peaks at $\vrel \approx 35 \vth$. In general, there are no ions at the velocity where $\sigma$ is maximized (the Maxwellian is $\sim 10^{-136}$ smaller here than in the thermal range). Only in a narrow window around the Gamow velocity ${v_* = b^{1/3}\vth/\sqrt{2}}$ are the distribution function and the fusion cross section simultaneously large enough to contribute significantly to the reactivity.

\subsection{Reactivity of perturbed distributions}

Of particular importance in this work is the fact that, as a result of the peaked reactivity integrand, $\avg{\sigma v}$ is extremely sensitive to perturbations of $f$ within the Gamow window. Consider a transformation 
\begin{equation}
  \label{eq_f_pert_g}
  f(\bs p) \to g (\bs p) f(\bs p),
\end{equation}
where $g$ varies slowly relative to $\fm$ and $\sigma$. The modification factor $g$ may, for example, be the result of wave-particle interactions that rearrange elements of $f$ in velocity space. Particularly in a controlled fusion device, it is of interest to calculate the change in reactivity resulting from such an interaction. Let $\Phi$ be the reactivity enhancement factor such that
\begin{equation}
  \label{eq_Phi_defn}
  \Phi = \frac{\avg{\sigma v}}{\avg{\sigma v}_0} ,
\end{equation}
where $\avg{\sigma v}_0$ is the unperturbed reactivity, and $\avg{\sigma v}$ is the reactivity of the distribution perturbed according to \eqref{eq_f_pert_g}. 
Starting from \eqref{eq_reac_def}, the reactivity integral then takes a form identical to \eqref{eq_reac_maxwellian} except for an additional factor $\varphi(\prel,\ptot)$ in the integrand, where 
\begin{equation}
  \varphi(\bs \prel,\bs \ptot) = g\paren{\frac{\bs \ptot - \bs \prel}{\sqrt{2}}} g \paren{\frac{\bs \ptot + \bs \prel}{\sqrt{2}}} .
\end{equation}
By construction, the integrand is still dominated by the rapidly varying exponential term. To compute the integral over $\prel$, one can therefore simply evaluate $\varphi$ at the Gamow peak, where $\prel = \sqrt{2}p_*$. Assuming that $g$ can be approximated by a power law $g \propto p^h$ in the Gamow window (and $h \ll b^{1/3}$), expanding in $\ptot$ about the Gamow peak shows that the integral over $\ptot$ is unchanged by the inclusion of the $\varphi$ factor to leading order in $b \gg 1$. The result is that 
\begin{equation}
  \label{eq_Phi_peak}
  \Phi \sim \avg{\varphi(b^{1/3} \bs {\hat r},0)}_\Omega = \left\langle g\paren{\frac{b^{1/3}}{\sqrt{2}} \bs {\hat r}} g\paren{-\frac{b^{1/3}}{\sqrt{2}} \bs {\hat r}}\right\rangle_\Omega ,
\end{equation}
up to corrections of order $b^{-2/3}$, where $\avg{\cdot}_\Omega$ indicates an average over the unit sphere and $\bs {\hat r}$ is a unit vector indicating directions over which the average is taken.

The takeaway of \eqref{eq_Phi_peak} is that, for modest perturbations to a near-Maxwellian distribution, only the region around the Gamow peak is significant for the reactivity. 
Notably, this region makes a much smaller contribution to the bulk properties of the plasma, such as fluid moments and fluxes. To illustrate this, consider a Maxwellian perturbed -- perhaps an isotropized beam of fast ions -- by a multiplicative correction described by a narrow Gaussian of width $\Delta$ centered at the Gamow peak $p_* = b^{1/3}/\sqrt{2}$, \textit{viz.} 
\begin{equation}
  \label{eq_example_pert_g}
  g(\bs p) = 1 + g_0 \e{-\half \paren{p - b^{1/3}/\sqrt{2}}^2/\Delta^2}
\end{equation}
($g_0$ is a constant). In the asymptotic limit $\Delta \ll 1 \ll b^{1/3}$, the reactivity can be approximated by averaging the beam distribution over the Gamow window and then applying \eqref{eq_Phi_peak}; the enhancement factor is
\begin{equation}
  \label{eq_Phi_example_beam} 
  \Phi \sim 1 + \sqrt{6} g_0 \Delta ,
\end{equation} 
assuming that $g_0 \lesssim 1$. 
The number density and energy density of the perturbation are proportional to the normalized perturbed moments $M_0$ and $M_2$, respectively, defined by
\begin{equation}
  \label{eq_M_j_defn}
  M_j = \frac{\sqrt{2\pi}}{\Gamma\paren{\frac{j+3}{2}}2^{\frac{j+3}{2}}}\int d^3 p p^j (g(\bs p) - 1) f(\bs p) ,
\end{equation}
where $\Gamma$ is the Gamma function. Using the perturbation in \eqref{eq_example_pert_g} and taking ${\Delta \ll b^{-1/3}}$ for simplicity, we have
\begin{equation}
  \label{eq_M0_example}
  M_0 \sim g_0 \Delta b^{2/3} \e{-\frac{1}{4}b^{2/3}}
\end{equation}
and 
\begin{equation}
  \label{eq_M2_example}
  M_2 \sim g_0 \frac{\Delta}{6} b^{4/3} \e{-\frac{1}{4}b^{2/3}} .
\end{equation}

\begin{figure}
    \centering 
    \includegraphics[width=0.95\columnwidth]{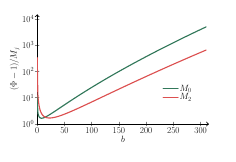}
    \caption{\justifying Normalized increase in reactivity $\Phi - 1$ compared to normalized change in moments $M_j$ of the distribution function as a function of the barrier-penetration parameter $b$. The perturbation \eqref{eq_example_pert_g} is taken to be a narrow beam centered at the Gamow peak.}
    \label{fig_M_j_norm}
\end{figure}

Evidently, perturbations to moments of the distribution function are exponentially small in $b$; as the Gamow peak moves further out onto the tail, the number of particles in the Gamow window becomes smaller relative to the number of thermal particles. Hence, multiplicative perturbations near $p_*$ become increasingly insignificant relative to the shape of the distribution in the thermal bulk. This is illustrated in Fig.~\ref{fig_M_j_norm}, where the normalized reactivity enhancement given in \eqref{eq_Phi_example_beam} is shown divided by the perturbed moments given in \eqref{eq_M0_example} and \eqref{eq_M2_example}. While the total enhancement, and the perturbed moments, depend on the height $g_0$ and width $\Delta$ of the perturbation, the important point is that the \textit{reactivity enhancement per unit change in the fluid moments} increases by orders of magnitude as $b$ becomes large. In other words, even perturbations producing sub-percent level changes in the density and temperature can modify the reactivity by an order-unity factor.

The reactivity enhancement considered in \eqref{eq_Phi_example_beam} is intrinsically non-thermal, meaning here that it is much larger than the enhancement obtained from a thermal distribution with the same moments. 
This is generically true of kinetic perturbations localized around the Gamow peak; if such a perturbation were to thermalize -- i.e. to redistribute its particles and energy into a Maxwellian of higher density and temperature than the original distribution -- most of the redistributed energy would end up in regions far from the Gamow peak and so would no longer contribute significantly to the reactivity. Suppose that the unperturbed distribution is a Maxwellian with density $n_0$ and temperature $T_0$. When the perturbation thermalizes, the distribution reaches a new equilibrium with density and temperature $n_\mr{eq}$ and $T_\mr{eq}$ given, to leading order in $b$, by 
\begin{equation}
  n_\mr{eq} = n_0(1 + M_0) \qquad \text{and} \qquad T_\mr{eq} = T_0(1 + M_2) .
\end{equation}
The equilibrium Gamow parameter is ${b_\mr{eq} = b(1 - \half M_2)}$. Assuming that the reactivity can be approximated by \eqref{eq_reac_maxwellian_approx}, it follows that after thermalization, the reactivity enhancement factor ${\Phi_\mr{eq} = \avg{\sigma v}_{T=T_\mr{eq}}/\avg{\sigma v}_{T=T_0}}$ is
\begin{equation}
  \label{eq_Phi_therm_example_beam}
  \Phi_\mr{eq} \sim 1 + g_0 \frac{\Delta}{12}b^2 \e{-\frac{1}{4}b^{2/3}} .
\end{equation}
Comparison of \eqref{eq_Phi_example_beam} and \eqref{eq_Phi_therm_example_beam} illustrates the asymptotically large difference between the kinetic reactivity enhancement and the thermal enhancement attained with the same amount of energy. 

In summary, when the temperature is far below the Gamow energy ($b \gg 1$), fusion reactivity and fluid moments are governed by distinct, and widely separated, regions of velocity space (the Gamow peak and the thermal bulk, respectively). In this limit, a given perturbation may affect one of these classes of quantities more than the other by an asymptotically large factor. The behavior illustrated in Fig.~\ref{fig_M_j_norm} for density and temperature also applies, for example, to momentum, heat flux, and viscous stress. 
The separation in velocity space has another important, although somewhat less obvious, consequence: because collision frequencies in a plasma are sensitive to velocity, the collisional dynamics of ions in the Gamow window can be very different from those of thermal particles. This distinction, which is critical to the SFRE, is the subject of the following section. 

\subsection{Coulomb collisions}

The utility of non-thermal distributions, including for increasing reactivity, is limited by their tendency to relax rapidly to equilibrium. In a Maxwellian (or nearly Maxwellian), classical, weakly coupled plasma, the characteristic frequency $\nu_{\sone\stwo}$ of collisions of species $\sone$ with species $\stwo$ is 
\begin{equation}
  \label{eq_nu_sone_stwo}
  \nu_{\sone\stwo} = \frac{4\sqrt{\pi}}{3} \frac{n_\beta Z_\sone^2 Z_\stwo^2 e^4 \sqrt{2\mu_{\sone \stwo}}\ln \Lambda}{ m_\sone T^{3/2}},
\end{equation}
where $Z_\sone$ and $Z_\stwo$ are the ion charge states, $e$ is the elementary charge, $m_\sone$ and $m_\stwo$ are the ion masses, ${\mu_{\sone\stwo} = m_\sone m_\stwo/(m_\sone + m_\stwo)}$ is the reduced mass, $n_\sone$ and $n_\stwo$ are the ion number densities, $T$ is the temperature, and $\ln\Lambda$ is the Coulomb logarithm. 
The effective collision frequency for each species is 
\begin{equation}
  \label{eq_nu_0}
  \nu_\sone = \sum_\stwo \nu_{\sone\stwo} ,
\end{equation}
and the collision time is ${\tau_{\sone\stwo} = 1/\nu_{\sone\stwo}}$. In general, $\nu_{\sone\stwo}$ is much larger than the fusion rate per ion $n\avg{\sigma v}$. Hence, while a non-thermal perturbation can briefly increase the fusion rate by a substantial factor, only a small fraction of ions will fuse before collisions destroy the perturbation.

\begin{comment}
The collision operator $\mc C$ can be written as
\begin{equation}
  \label{eq_C_general}
  \mc C[f] =     ,
\end{equation} 
where $\widehat \nu_s (p)$, $\widehat \nu_\parallel(p)$, and $\widehat \nu_\perp(p)$ are dimensionless functions of velocity describing rates of slowing, diffusion to higher or lower velocity, and pitch-angle scattering, normalized to $\nuii$. 
\end{comment}
Missing from this picture, however, is the fact that collision frequency varies rapidly with velocity. 
For fast ions, the collisions with other ions are dominated by slowing down and pitch-angle scattering, the rates of which scale asymptotically as $p^{-3}$ when $p \gg 1$. We can therefore define an approximate ``effective collision frequency'' $\widehat \nu$ as
\begin{equation}
  \label{eq_nu_p}
  \widehat \nu(p) = \frac{1}{1 + p^3} + \frac{1}{Z}\sqrt{\frac{m_e}{m_\sone}} ,
\end{equation}
where $m_e$ is the electron mass. The second term accounts for slowing down by collisions with electrons; the effect of electron slowing is small at thermal velocities because the electron-ion mass ratio is small, but it becomes important at sufficiently high velocities.

\begin{figure}
    \centering 
    \includegraphics[width=0.95\columnwidth]{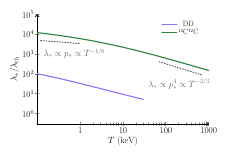}
    \caption{\justifying Ratio of the Gamow mean free path $\lambda_*$ to the thermal mean free path $\lth$ as a function of temperature for DD and \textsuperscript{12}C\textsuperscript{12}C reactions.} 
    \label{fig_lambda_star_ratio}
\end{figure}

The strong dependence of collision frequency on velocity is a notable feature distinguishing plasmas from other fluids (in a gas of hard spheres, for example, $\widehat \nu \propto p$). A consequence of this dependence is that the fast ions responsible for fusion reactions are much less collisional than their thermal counterparts. Let the ``Gamow collision frequency'' $\nu_* = \nu_\sone \widehat \nu(p_*)$ be the collision frequency of ions at the Gamow peak. %, and let $\tau_* = 1/\nu_*$ be the corresponding collision time. 
It is useful also to define a ``Gamow mean free path'' $\lambda_* = v_*/\nu_*$ as the distance traveled between collisions by an ion at the Gamow peak, so
\begin{equation}
  \label{eq_lambda_star}
  \lambda_* = \frac{\lth 2^{-1/2}b^{1/3}}{\frac{1}{1 + 2^{-3/2}b} + \frac{1}{Z_\sone}\sqrt{\frac{m_e}{m_\sone}}} . 
\end{equation}
In the limit $b \gg 1$, $\lambda_*$ becomes asymptotically larger than the thermal mean free path $\lth = \vth/\nu_\sone$. The scaling of $\lambda_*/\lth$ with temperature (and hence with $b$) is illustrated in Fig.~\ref{fig_lambda_star_ratio} for the DD and \textsuperscript{12}C\textsuperscript{12}C reactions. In the low-temperature limit, where particles near the Gamow peak slow primarily on electrons, ${\lambda_*/\lth \propto p_* \propto b^{1/3} \propto T^{-1/6}}$. At higher temperatures, ion slowing dominates and, provided that $b \gg 1$ is still satisfied, ${\lambda_*/\lth \propto p_*^4 \propto b^{4/3} \propto T^{-2/3}}$. In either case, the Gamow mean free path can be orders of magnitude larger than the thermal mean free path. The dramatic separation between thermal and fusion-relevant scales is central to the kinetic physics described in this work.

It bears mentioning that the particular scaling of the collision frequency assumed in \eqref{eq_nu_p}, and the resulting form of the Gamow mean free path in \eqref{eq_lambda_star}, is not critical to the physics described in this work. 
The key point is that fast ions travel much further between collisions than thermal ions do, making the fast-ion population more sensitive to gradients in plasma properties. 
Already, the form of the collision frequency in \eqref{eq_nu_p} is a reduced model, adopted in the interest of analytical simplicity, that captures the qualitative behavior of collisions in a weakly coupled, classical plasma. 
In reality, the stopping power of fast ions can be modified by effects including large-angle scattering \cite{Li_Petrasso_1993a}, collective effects \cite{Li_Petrasso_1993b}, and electron degeneracy \cite{Son_Fisch_2005}. 
Non-ideal stopping powers are of particular interest in the transport and energy deposition profiles of alpha particles\cite{Reichelt_Petrasso_Li_2024,Du_Kang_Zou_Liu_Deng_Ge_Dai_Cai_Zhu_2024,Zylstra_Hurricane_2019}, and of fast ions (which may be used, for example, in some fast-ignition schemes) \cite{Malko_et_2022}, but the stopping powers of fast ions in the Gamow window may also be modified in some regimes. 
However, the general behavior of the fast-ion collision rate captured in \eqref{eq_nu_p} is expected to obtain in nearly all laboratory and astrophysical systems of interest. 
The asymptotic formulas derived subsequently in this work, for instance \eqref{eq_G_k_two_spec}, report results in terms of the Gamow mean free path and do not require assuming a specific form of the collision frequency, aside from the property that $\lambda_*/\lth \gg 1$. 
Hence, while quantitative refinements in the collision model may be called for in specific systems, either to account for non-ideal stopping-power effects or to correct for deficiencies of the collision operator described in \S\ref{sec_turb_reac}, the asymptotic and qualitative results of this work are expected to be robust with respect to such refinements.

\begin{comment}
In both analytical calculation and numerical simulation, the collision operator described in \eqref{eq_C_general} is often prohibitively complicated. In some cases, a much simpler Bhatnagar-Gross-Krook (BGK) operator is an adequate approximation, taking the form 
\begin{equation}
  \label{eq_C_BGK}
  \mc C[f] = -\nu (f - \fm) ,
\end{equation}
where $\fm$ is a Maxwellian matching the zeroth, first, and second moments of $f$, and $\nu$ is an appropriately chosen collision frequency.
\end{comment}

\section{Shear flow reactivity enhancement}
\label{sec_sfre}

Taken together, the physical scalings described in \S\ref{sec_background} lead to surprising effects in plasmas whose fluid properties vary in space. 
Remarkably, in the presence of flows sheared on length scales comparable to the Gamow mean free path, fusion reactivity can increase several-fold relative to a plasma with the same thermal energy. This section offers a largely qualitative physical model of this ``shear flow reactivity enhancement effect'' (SFRE)\cite{Fetsch_Fisch_2025a}. A more detailed and quantitative treatment follows in \S\ref{sec_turb_reac}. 
In the spirit of the model adopted in Ref.~\citenum{Fetsch_Fisch_2025a}, the SFRE can be understood as follows.

\begin{figure}
    \centering 
    \includegraphics[width=0.95\columnwidth]{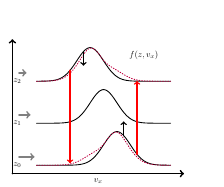}
    \caption{\justifying Distribution functions (black) in a planar shear flow. Arrows show the ballistic transport of thermal particles (black) and fast particles (red) in a sheared flow. The longer mean free paths of fast particles allow them more easily to cross the flow gradient and enhance the tail. Figure based on Ref.~\citenum{Fetsch_Fisch_2025a}.} 
    \label{fig_sfre_cartoon}
\end{figure}

Consider, as sketched in Fig.~\ref{fig_sfre_cartoon}, a plasma containing a planar shear flow $\bs u = u(z) \bs e_x$, where the flow velocity is in the $x$ direction ($\bs e_x$ is the x-directed unit vector), and the flow magnitude varies on a length scale $L$ in the $z$ direction. 
Assuming that no other dynamics are driving the distribution out of local thermal equilibrium, any kinetic deviations from an equilibrium Maxwellian distribution will be sourced by the flow shear. A conventional heuristic for the size of these kinetic corrections is the Knudsen number
\begin{equation}
  \label{eq_Kn_sfre}
  \mr{Kn} = \frac{\lambda_\mr{th}}{L} .
\end{equation}
When $\mr{Kn} \ll 1$, particles traveling across a flow gradient are unable to sample large variations in the background flow between collisions, and so the distribution function remains close to a Maxwellian. 
This heuristic can break down, however, for particles in the tail of the distribution. Following  Refs.~\citenum{McDevitt_Tang_Guo_2017,Fetsch_Fisch_2025a}, we define the Gamow-Knudsen number $\mr{Gk}$ to be 
\begin{equation}
  \label{eq_Gk_sfre}
  \mr{Gk} = \frac{\lambda_*}{L} ,
\end{equation}
noting that $\mr{Gk}/\mr{Kn} \sim \mc O(b^{4/3})$ in the ion-stopping regime (cf. Fig.~\ref{fig_lambda_star_ratio}). Hence, when $b \gg 1$, the condition $\mr{Kn} \ll 1$ does not guarantee that the distribution function in the Gamow window is close to Maxwellian. In fact, even modest flows can generate large deviations from Maxwellian in the tail, leading to substantial reactivity enhancements.

To see how flow shear affects the tail of the ion distribution, consider, as in Fig.~\ref{fig_sfre_cartoon}, an ion sampled from a Maxwellian at a point $z_0$ with velocity $\bs v$, relative to the local flow, and mean free path $\lambda$. If this ion happens to have a velocity component in the $z$ direction, it will travel ballistically some distance $\lambda v_z/v$ across the flow gradient before experiencing a collision. If $\lambda$ is sufficiently large compared to $L$, the ion will reach a different point $z_1$ where the flow velocity is different. If $\bs v \cdot (\bs u(z_1) - \bs u(z_0))$ is negative, then the peculiar velocity of the ion relative to the local flow at $z_1$ is larger than was its peculiar velocity at its starting location $z_0$. 
If another ion is sampled at $z_0$ with speed $v' > v$, its mean free path $\lambda'$ is longer than that of the first ion. If the orientations of $\bs v'$ and $\bs v$ are similar, then the second ion will travel further across the flow gradient, reaching, perhaps, a point $z_2$ where the flow differential compared to $z_0$ is larger than at $z_1$. This second ion thus has a larger peculiar velocity than that of the first ion.

The fact that ions arrive at $z_1$ and $z_2$ with increased peculiar velocities means that the distribution function at these points must have an enhanced tail; the size of the enhancement must increase with velocity due to the increase in mean free path. When the tail is enhanced in the vicinity of the Gamow peak, more ions are able to fuse, leading to an increase in reactivity; this is the SFRE.

Because particles leave $z_0$ to reach $z_1$ and $z_2$, it is reasonable to ask whether the distribution at $z_0$ is depleted in the tail. In fact, the enhancement of the tail dominates over any depletion. To illustrate this point, let the ion phase space be projected onto the $z$-$v_x$ plane, and consider a point ${\bs q = (z_1, w_x)}$ where ${w_x > 0}$. Ions arriving from $z_0$ with $v_z > 0$, and those arriving from $z_2$ with $v_z < 0$, enhance the distribution at $\bs q$, while ions leaving from $z_1$ in either direction deplete the distribution. 
As a simple model of the process, let $\phi$ be the fraction of ions of velocity $v \sim w$ that are able to cross a distance $L$ per unit time without colliding along the way. If the distributions start as exactly Maxwellian everywhere, then the initial time evolution of the distribution at $z_1$ can be estimated as
\begin{equation}  
  \label{eq_f_z1_evol_sfre}
  \frac{\partial f(z_1, \bs w)}{\partial t} \sim \frac{\phi}{2} f(z_0, \bs w)  + \frac{\phi}{2} f(z_2, \bs w) - \phi f(z_1, \bs w) ,
\end{equation}
where the factors of $1/2$ account for the fact that only half of the ions at $z_0$ and $z_2$ are traveling in the right direction to reach $z_1$. 

\begin{figure}[t]
  \centering
  \includegraphics[width=0.95\columnwidth]{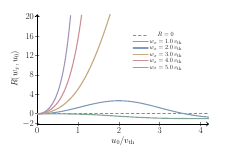}
  \caption{\justifying Behavior of $R(w_x,u_0)$ as a function of $u_0/v_\mathrm{th}$ for representative values of $w_x/v_\mathrm{th}$. For suprathermal $w_x$ and small to moderate flow amplitudes, $R$ is positive, indicating that the number of particles on the tail increases.}
  \label{fig_R_vx_u0}
\end{figure}

The key point is that, in \eqref{eq_f_z1_evol_sfre}, the ions arriving from $z_0$ are coming from a region closer to the peak of the Maxwellian than at $z_1$. Ions arriving from $z_2$ are coming from a region further from the peak. Because of the exponential fall-off of $\fm$ with velocity, this leads to a larger number of particles arriving at $(z_1, w_x)$ than are leaving it. (The asymmetric particle flux is compensated by depletion of some lower-velocity regions, but these regions have a small effect on reactivity). 
Let ${z_2 - z_1 = z_1 - z_0 = L}$, and let ${u(z) = -u_0 z/L}$, where $u_0$ is a characteristic velocity. Then \eqref{eq_f_z1_evol_sfre} becomes
\begin{equation}
  \label{eq_f_z1_evol_sfre_2}
  \frac{\partial f(z_1, \bs w)}{\partial t} \sim \frac{\phi\e{-\half w^2/\vth^2}}{2(2\pi)^{3/2}\vth^3} \left[ \e{(w_x u_0 - \half u_0^2)/\vth^2} + \e{(-w_x u_0 - \half u_0^2)/\vth^2} - 2 \right] ,
\end{equation}
meaning that 
\begin{equation}
  \label{eq_f_z1_evol_sfre_3}
  \frac{\partial f(z_1, \bs w)}{\partial t} \propto R(w_x, u_0) = \cosh\paren{\frac{w_x u_0}{\vth^2}}\e{-\half u_0^2/\vth^2} - 1 ,
\end{equation}
where the function $R(w_x,u_0)$ is plotted in Fig.~\ref{fig_R_vx_u0}.
For ions with suprathermal velocities in the $x$ direction (${w_x > \vth}$), and for flows that are not significantly supersonic (${u_0 \lesssim \vth}$), \eqref{eq_f_z1_evol_sfre_3} guarantees that $\partial f(z_1, w_x)/\partial t > 0$, meaning that the effect of transport on a near-Maxwellian background is to enhance the tail of the distribution at $z_1$.
While the model described here is heuristic, the qualitative conclusion is substantiated by numerical simulations in Ref.~\citenum{Fetsch_Fisch_2025a,Fetsch_Fisch_2025b} and by the kinetic theory presented in the following section. Moreover, the tendency of the distribution to spread out into an enhanced tail is expected on entropic grounds.

As is apparent from Fig.~\ref{fig_sfre_cartoon}, a larger flow velocity leads to a larger reactivity enhancement. Interestingly, however, the SFRE can be significant (i.e. of order unity or greater) even when the flow velocity is small compared to the thermal velocity. Because the Maxwellian distribution falls off exponentially in $v^2$, even a small upshift in the peculiar velocity of some ions yields an exponentially large increase in the number of ions in the Gamow window, provided that the Gamow peak is far out on the tail. Using the value of $R(w_x, u_0)$ as a heuristic for enhancement factor in the tail, and hence for the size of the SFRE, this exponential scaling can be seen in Fig.~\ref{fig_R_vx_u0}, where even subsonic flows can produce an order-of-magnitude increase in the tail population.

\section{Turbulence and reactivity}
\label{sec_turb_reac}

The previous section considered a simple planar shear flow in the interest of giving a transparent description of the physics behind the SFRE. Extending the theory to more general cases -- for instance, computing the reactivity enhancement in a disordered, three-dimensional turbulent plasma -- requires a more systematic quantitative theory. The derivation of such a theory is the subject of this section.

\subsection{Setup}

Consider an unmagnetized, classical, weakly coupled plasma consisting of one or more ion species indexed by $\sone$, as well as electrons. Let each ion species have mass $m_\sone$, charge state $Z_\sone$, and number density $n_\sone$. The electron mass is $m_e$, and the electron number density is $n_e = \sum_\sone Z_\sone n_\sone$. The ion species are assumed to have the same temperature $T$, which is spatially uniform. The flow velocity $\bs u(\bs x)$ varies with position $\bs x$ but is equal for all species at each point. Flows are assumed to be incompressible  ($\nabla \cdot \bs u = 0$). 
As in \S\ref{sec_background}, let $\nu_\sone$ be the frequency of collisions between thermal ions of species $\sone$ and all other ions ($\tau_\sone = 1/\nu_\sone$), and let $\lambda_{\sone,\mr{th}} = \vthone/\nu_\sone$ be the thermal mean free path of each species. We will assume that the Knudsen number of the flow is small, meaning that 
\begin{equation}
  \label{eq_Kn_cond_turb}
  \frac{\partial_i u_j}{|u|} \lambda_{\sone,\mr{th}} \ll 1 
\end{equation}
for all indices $i$ and $j$, and for all species $\sone$, at all points in the plasma (or, at least, that \eqref{eq_Kn_cond_turb} is violated in a negligible fraction of the region of interest). We do not require, however, that the Gamow-Knudsen number $\mr{Gk} \sim \mc O(b^{4/3} \mr{Kn})$ be small. 
For the sake of analytical tractability, we will further assume that the Mach number of the flow is small, meaning that 
\begin{equation}
  \label{eq_Ma_cond_turb}
  \frac{|u|}{\vthone} \ll 1 ,
\end{equation}
for all ion species and at all points. 

When the Knudsen number and Mach number are sufficiently small, kinetic calculations can be performed while treating the flow as stationary in time. Noting that the characteristic turnover time $\tau_L$ of an eddy of scale $L$ and speed $u$ is $\tau_L \sim L/u$, we have
\begin{equation}
  \label{eq_tau_L_tau_i}
  \frac{\tau_L}{\tau_\sone} \sim \mc O\paren{\mr{Kn}^{-1} \mr{Ma}^{-1}} \gg 1 ,
\end{equation}
so an ion experiences many collisions within the time required for the flow to rearrange itself significantly. The same ordering allows dissipation to be neglected. The kinematic viscosity $\eta$ can be written $\eta \sim c_\eta \lambda_{\sone,\mr{th}} \vthone$, where $c_\eta$ is a constant of order unity, and the dissipation time $\tau_\eta$ for eddies of scale $L$ is $\tau_\eta \sim L^2/2\eta$. Because
\begin{equation}
  \label{eq_tau_eta_tau_i}
  \frac{\tau_\eta}{\tau_\sone} \sim \mc O\paren{\mr{Kn}^{-2}} \gg 1 ,
\end{equation}
ions experience many collisions before the flow relaxes significantly due to viscosity. 
Effectively, then, the ion distribution functions evolve in a frozen fluid background.

Finally, for the sake of a closed-form analytical treatment, we will approximate collisions, as in \S\ref{sec_sfre}, by a BGK operator, \textit{viz.}
\begin{equation}
  \label{eq_BGK_sone}
  \mc C_\sone[f_\sone] = -\nu_\sone \widehat \nu_\sone(p_\sone) (f_\sone - \fm) ,
\end{equation}
where $\bs p_\sone = (\bs v - \bs u)/\vthone$ is the normalized peculiar velocity for species $\sone$, and $\widehat \nu_\sone$ is the dimensionless coefficient describing the velocity dependence of collisions between $\sone$ and all other species, normalized to $\nu_\sone$. 
When $\widehat \nu_\sone$ is not a constant, \eqref{eq_BGK_sone} does not, in general, conserve particles, momentum, or energy. The operator is therefore unsuitable for precise calculation of fluid moments or transport coefficients. For the SFRE, however, due to the asymptotic scale separation outlined in \S\ref{sec_background}, we are concerned only with the tail of the distribution function; \eqref{eq_BGK_sone} offers a tractable model capturing the velocity dependence of particle transport in this region\cite{Fetsch_Fisch_2025a,Fetsch_Fisch_2025b}.

\subsection{Kinetic perturbations}

By the mechanism discussed in \S\ref{sec_sfre}, spatial variations in $\bs u(\bs x)$ produce non-thermal perturbations to the ion distribution functions. In the absence of external forces, and adopting the assumption of a ``frozen background'' substantiated by \eqref{eq_tau_L_tau_i} and \eqref{eq_tau_eta_tau_i}, the steady-state distribution function $f_\sone$ can be computed by solving
\begin{equation}
  \label{eq_kinetic_eq_steady}
  \paren{\bs p + \frac{\bs u}{\vthone}} \cdot \nabla f_\sone - \frac{\bs p \cdot \nabla \bs u}{\vthone} \cdot \frac{\partial f_\sone}{\partial \bs p} = -\frac{\widehat \nu_\sone(p_\sone)}{\lthsone} (f_\sone - \fm) .
\end{equation}
To work in Fourier space, we define the normalized, Fourier-transformed flow field $\wt {\bs u}(\bs k)$ as 
\begin{equation}
  \label{eq_u_fourier}
  \wt {\bs u}_\sone(\bs k) = \frac{1}{\vthone V} \int d^3 x \e{-i \bs k \cdot \bs x} \bs u(\bs x),
\end{equation}
where $V$ is the volume of the system. As in Ref.~\citenum{Fetsch_Fisch_2025b}, we expand \eqref{eq_kinetic_eq_steady} in the Mach number, \textit{viz.}
\begin{equation}
  \label{eq_Mach_expansion}
  f_\sone = f_{\sone,0} + f_{\sone,1} + f_{\sone,2} + \cdots ,
\end{equation}
and solve for $f_\sone$ order by order. The leading-order solution is $f_{\sone,0} = \fm$. At first order, the contribution from each mode $\bs k$ of the flow is independent, so the Fourier-transformed first-order distribution $\wt f_{\sone,1}(\bs k, \bs p)$ corresponding to each mode is given by
\begin{equation}
  \label{eq_kinetic_eq_f1}
  i \bs p\cdot \bs k \wt f_{\sone,1} - i \bs p \cdot \bs k \wt {\bs u}_\sone \cdot \frac{\partial \fm}{\partial \bs p} = -\frac{\widehat \nu_\sone(p_\sone)}{\lthsone} \wt f_{\sone,1} .
\end{equation}
It is useful to define the quantity $\gamma(k, p)$ as
\begin{equation}
  \label{eq_gamma_defn}
  \gamma_\sone(k, p) = \frac{k p \lthsone}{\widehat \nu_\sone(p_\sone)} ,
\end{equation}
noting that $\gamma_\sone \sim \mc O(\mr{Gk})$. 
For each $k$, we can define, without loss of generality, a coordinate system such that $\bs k = k \bs e_z$ and $\wt {\bs u}_\sone = \wt u \bs e_x$, where $\bs e_x$ and $\bs e_z$ are unit vectors in the $x$ and $z$ directions, respectively. In this coordinate system, the solution to \eqref{eq_kinetic_eq_f1} is
\begin{equation}
  \label{eq_f1}
  \wt f_{\sone,1}(\bs k, \bs p) = \wt u \frac{-i \gamma_\sone \xi \chi p}{1 + i\gamma_\sone \xi} \fm ,
\end{equation}
where $\xi = \bs p \cdot \bs e_z/p$ and $\chi = \bs p \cdot \bs e_x/p$. 

The second-order distribution $\wt f_{\sone,2}$ satisfies
\begin{widetext}
\begin{equation}
  \label{eq_kinetic_eq_f2}
  i \bs p\cdot \bs k \wt f_{\sone,2}(\bs k, \bs p) - \frac{V}{(2\pi)^3} \iint d^3 k' d^3 k'' ~ i \bs p \cdot \bs k' \wt{\bs u}_\sone(\bs k') \cdot \frac{\partial \wt f_{\sone,1}(\bs k'',\bs p)}{\partial \bs p} \delta(\bs k - \bs k' - \bs k'')= -\frac{\widehat \nu_\sone(p_\sone)}{\lthsone} \wt f_{\sone,2}(\bs k, \bs p) .
\end{equation}
\end{widetext}
The ultimate quantity of interest in this calculation is the volume-averaged reactivity. Note that, when $\wt f_{\sone,2}$ is averaged over space, only the $k=0$ component survives. This component, in turn, is sourced only by pairs of modes with wavenumbers $\bs k$ and $-\bs k$. Because the flow field is real, $\wt {\bs u}_\sone(-\bs k) = \wt {\bs u}_\sone^*(\bs k)$. We denote the $k=0$ component of $\wt f_{\sone,2}$ sourced by perturbations of wavenumber $\pm \bs k$ by ${f_{\sone,2}(\bs p; \bs k) = \half \wt f_{\sone,2}(0, \bs p; \bs k)}$,
where ${\wt f_{\sone,2}(\bs k', \bs p; \bs k)}$ is the solution to \eqref{eq_kinetic_eq_f2} corresponding to a perturbation of wavenumber $\bs k$ and its complex conjugate. 
The factor of $\half$ is included to avoid double-counting when summing over all $\bs k$. Adopting the same coordinate system as in \eqref{eq_f1}, some manipulation of \eqref{eq_kinetic_eq_f2} yields\cite{Fetsch_Fisch_2025b}
\begin{equation}
  \label{eq_f2}
  f_{\sone,2}(\bs p;\bs k) = |\wt u|^2 f_M \frac{\gamma_\sone^2 \xi^2}{1 + \gamma_\sone^2 \xi^2} \left[-1 + p^2 \chi^2 - g \frac{\chi^2(1 - \gamma_\sone^2\xi^2)}{1 + \gamma_\sone^2\xi^2}\right] ,
\end{equation}
where $g = \partial \ln \gamma_\sone/\partial \ln p$ is the logarithmic derivative of the Gamow-Knudsen number.

\begin{figure}[!hb]
  \centering
  \begin{subfigure}{\columnwidth}
    \centering
    \makebox[\columnwidth][c]{%
      \hspace*{\subfigpanelxshift}%
      \begin{tikzpicture}
        \node[inner sep=0] (imgA) {\includegraphics[width=0.95\columnwidth]{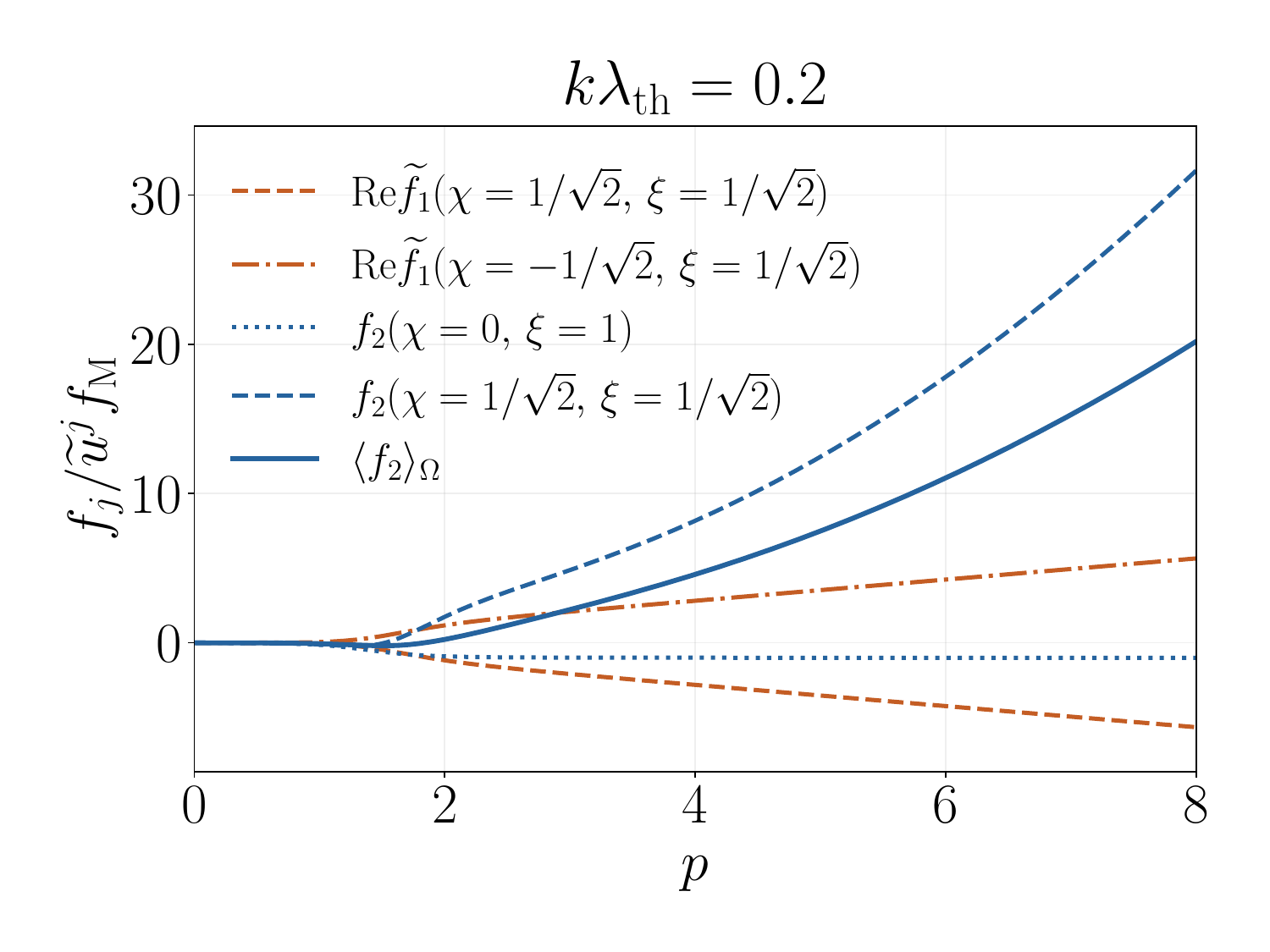}};
        \node[anchor=north west, font=\fontsize{\subfiglabelsizept pt}{\subfiglabelleadingpt pt}\selectfont] at ([xshift=0.075\linewidth,yshift=-0.02\linewidth]imgA.north west) {(a)};
      \end{tikzpicture}%
    }
  \end{subfigure}
  \vspace{0em}
  \begin{subfigure}{\columnwidth}
    \centering
    \makebox[\columnwidth][c]{%
      \hspace*{\subfigpanelxshift}%
      \begin{tikzpicture}
        \node[inner sep=0] (imgB) {\includegraphics[width=0.95\columnwidth]{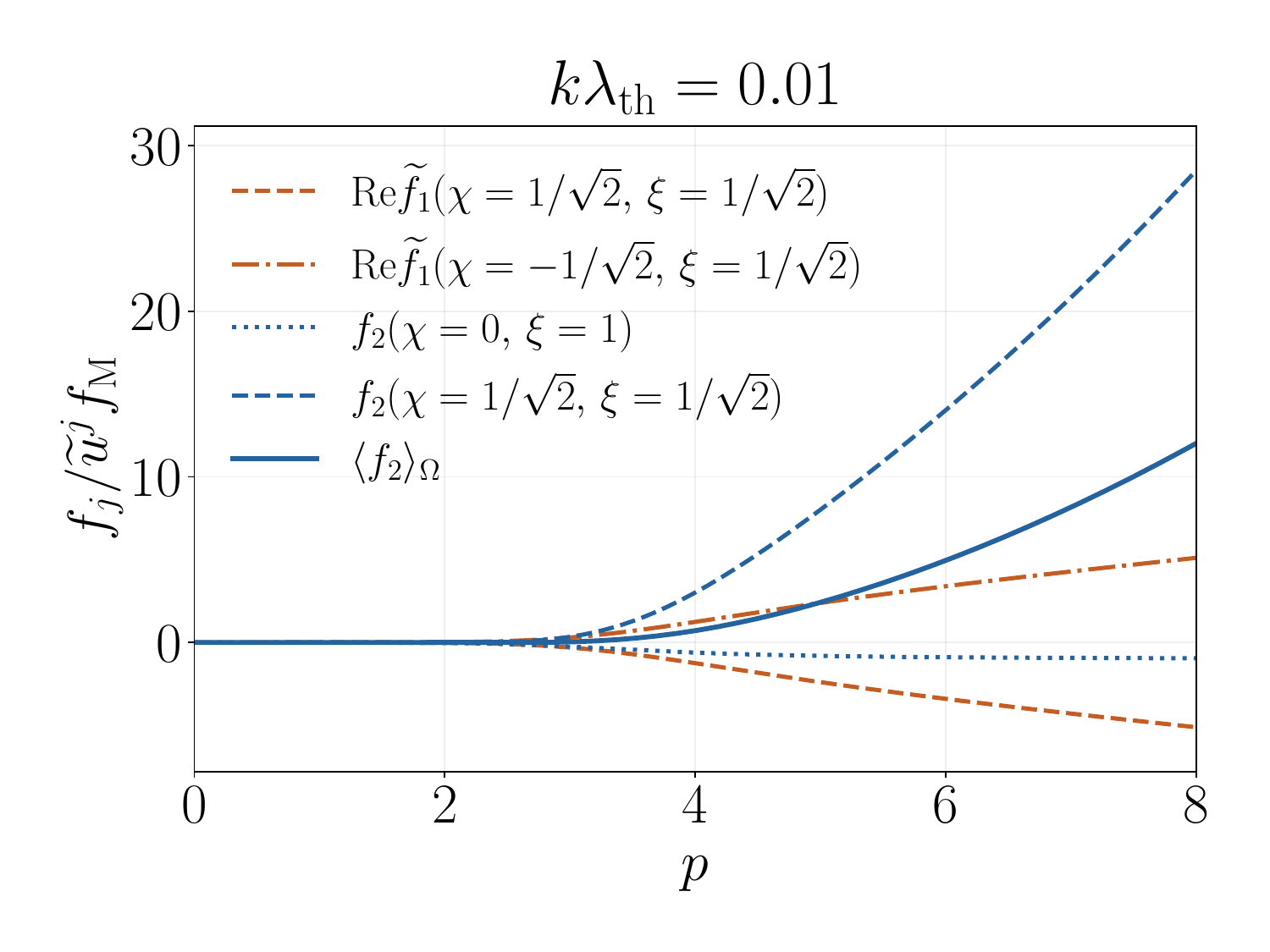}};
        \node[anchor=north west, font=\fontsize{\subfiglabelsizept pt}{\subfiglabelleadingpt pt}\selectfont] at ([xshift=0.075\linewidth,yshift=-0.02\linewidth]imgB.north west) {(b)};
      \end{tikzpicture}% 
    }
  \end{subfigure}
  \vspace{0em}
  \begin{subfigure}{\columnwidth}
    \centering
    \makebox[\columnwidth][c]{%
      \hspace*{\subfigpanelxshift+0.3cm}%
      \begin{tikzpicture}
        \node[inner sep=0] (imgC) {\includegraphics[width=0.95\columnwidth]{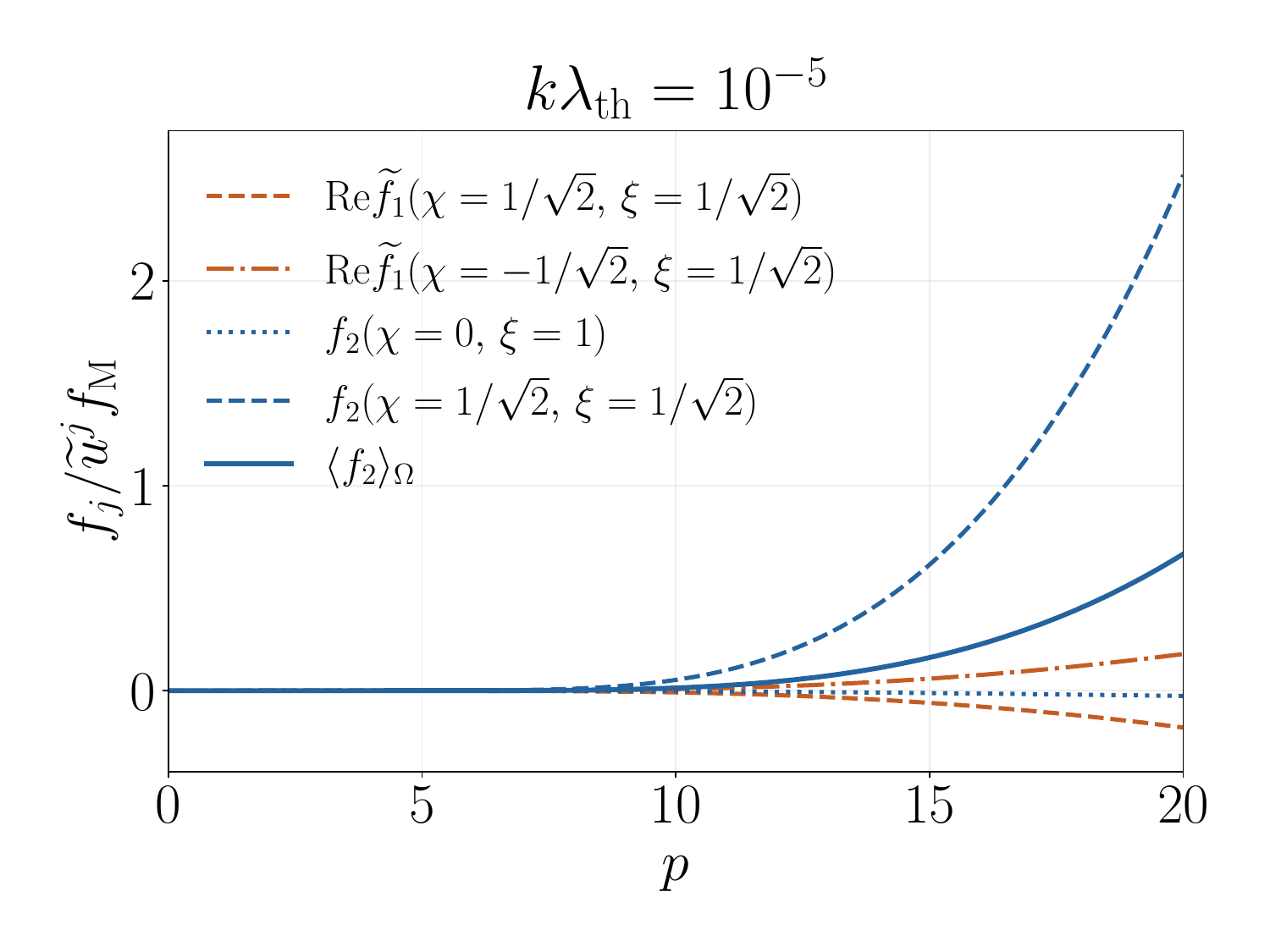}};
        \node[anchor=north west, font=\fontsize{\subfiglabelsizept pt}{\subfiglabelleadingpt pt}\selectfont] at ([xshift=0.075\linewidth,yshift=-0.02\linewidth]imgC.north west) {(c)};
      \end{tikzpicture}%
    }
  \end{subfigure}
  \caption{\justifying Perturbed distribution functions generated by flows with wavenumbers satisfying (a) $k\lth = 0.2$, (b) $k\lth = 0.01$, and (c) $k\lth = 10^{-5}$. The ion species is D in (a) and (b) and \textsuperscript{12}C in (c). A wider velocity range is shown in panel (c). The first-order perturbation $f_1$ is obtained from \eqref{eq_f1} and evaluated at $45^\circ$ and $135^\circ$ angles in the $p_x$-$p_z$ plane. The second-order perturbations $f_2$ and $\avg{f_2}_\Omega$ are obtained from \eqref{eq_f2} and \eqref{eq_f2_avg}.}
  \label{fig_f_k_pert} 
\end{figure}
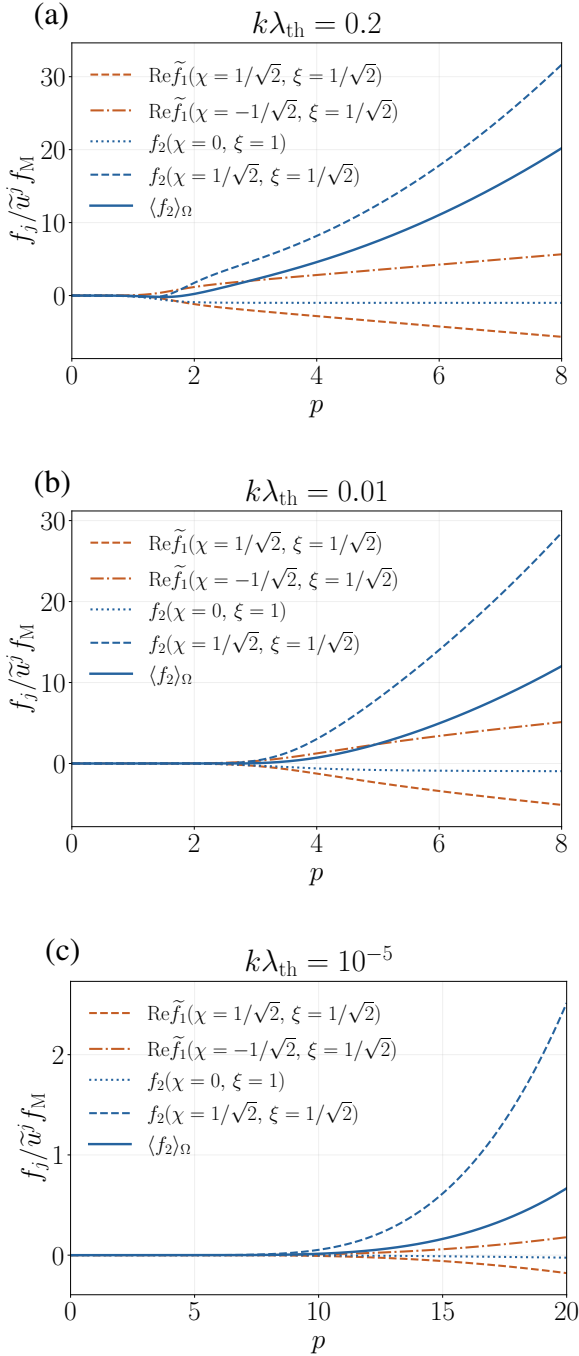

Finally, for reactions between the second-order perturbed distribution and a Maxwellian, the angular dependence of $\wt f_{\sone,2}$ is unimportant, so it is useful to compute the angular average $\avg{\wt f_{\sone,2}}_\Omega$ from \eqref{eq_f2}; following Ref.~\citenum{Fetsch_Fisch_2025b}, we find
\begin{equation}
  \label{eq_f2_avg}
  \begin{split}
  \left\langle f_{\sone,2}(\bs p; \bs k)\right\rangle_\Omega = &\frac{1}{4} |\wt u|^2 f_M \Bigg [ p^2 \frac{4\gamma_\sone^3 - 6(\gamma_\sone^2 + 1)\tan^{-1}(\gamma_\sone) + 6\gamma_\sone}{3\gamma_\sone^3} \\
  & - 4\paren{1 - \frac{\tan^{-1}(\gamma_\sone)}{\gamma_\sone}} \\
  & - g \frac{12(\gamma_\sone^2 + 2)\tan^{-1}(\gamma_\sone) - 4\gamma_\sone(\gamma_\sone^2 + 6)}{3\gamma_\sone^3}
  \Bigg ] .
  \end{split}
\end{equation}

The perturbed distributions are shown in Fig.~\ref{fig_f_k_pert} as functions of velocity. In panels (a) and (b), the ion species is deuterium, and in panel (c), it is carbon-12. (The ion species determines the relative rates of ion-ion and ion-electron collisions.) The first-order correction is complex, meaning that the perturbation to the distribution function is out of phase with the flow; the real part of $\wt f_1$ is shown in Fig.~\ref{fig_f_k_pert}. To illustrate the velocity-space anisotropy, slices are taken at $45^\circ$ and $135^\circ$ angles in the $p_x$-$p_z$ plane for $\wt f_1$ and at $0^\circ$ and $45^\circ$ for $f_2$. The perturbations are displayed as multiplicative corrections normalized to the background Maxwellian and to the flow, i.e. ${\wt f_1(\bs k, \bs p)/\wt u(\bs k) f_M}$ and ${f_2(\bs p; \bs k)/|\wt u(\bs k)|^2 f_M}$. 
Notably, the distributions remain close to Maxwellian within the thermal range before increasing rapidly in the tail. This feature, which is a consequence of the strong velocity dependence of the collision frequency, is critical to the SFRE, illustrating that sheared flows can generate large perturbations in the Gamow window without substantially distorting the thermal bulk. For flows on longer length scales (here, $k\lth = 0.01$ or $k\lth = 10^{-5}$ as opposed to $k\lth = 0.2$), the increase begins further out on the tail.

\subsection{Enhanced reactivity}

Using the expressions for $\avg{f_{\sone,2}}_\Omega$ and $\wt f_{\sone,1}$ in \eqref{eq_f2_avg} and \eqref{eq_f1}, along with \eqref{eq_Phi_peak}, it is straightforward to compute the reactivity enhancement resulting from an arbitrary subsonic flow field to second order in the Mach number. Following Ref.~\citenum{Fetsch_Fisch_2025b}, the enhancement factor $\Phi$ can be expressed as 
\begin{equation}
  \label{eq_Phi_E_G}
  \Phi \sim 1 + 2 \int_0^\infty dk E_\sone(k) G_\sone(k) ,
\end{equation}
where the normalized turbulent energy spectrum $E_\sone(k)$ for species $\sone$ is
\begin{equation}
  \label{eq_E_k_defn}
  E_\sone(k) = \half \frac{V}{(2\pi)^3} \int d^3 k' |\wt {\bs u}_\sone(\bs k')|^2 \delta(k - |\bs k'|) 
\end{equation}
and $G_\sone(k)$ is a dimensionless utility function describing the increase in reactivity per unit energy in the flow at wavenumber $k$. For a single reactant species, $G_\sone(k)$ is given in the limit $b \gg 1$ by\cite{Fetsch_Fisch_2025b}
\begin{equation}
  \label{eq_G_k_asymptotic}
  G_\sone(k) \sim \frac{b^{2/3}}{2} \paren{\frac{1}{3} + \frac{\tan^{-1}(\gamma_{\sone,*}) - \gamma_{\sone,*}}{\gamma_{\sone,*}^3}} ,
\end{equation}
where $\gamma_{\sone,*} = \gamma_\sone(k, b^{1/3}/\sqrt{2})$ is the Gamow-Knudsen number evaluated at the Gamow peak.

For reactions between different ion species $\sone$ and $\stwo$ with the same temperature $T$ and masses $m_\sone$ and $m_\stwo$, the reactivity operator defined in \eqref{eq_reac_def} can be written as
\begin{equation}
  \label{eq_reac_two_spec}
  \Sigma[f_\sone, f_\stwo] = A'\iint d^3 \ptot d^3 \prel \frac{S(\vrel)}{\prel} \frac{\varphi_{\sone \stwo}}{(2\pi)^3} \e{-b/\prel - \half \ptot^2 - \half \prel^2} ,
\end{equation}
where $A' = A\sqrt{\mu_{\sone\stwo}/T}$ and the center-of-mass coordinate system is now defined such that
\begin{equation}
  \label{eq_p_com_two_spec}
  \begin{split}
  \bs \ptot &= \sqrt{\frac{m_\sone}{M}} \bs p_\sone + \sqrt{\frac{m_\stwo}{M}} \bs p_\stwo , \\
  \bs \prel &= \sqrt{\frac{m_\stwo}{M}} \bs p_\sone - \sqrt{\frac{m_\sone}{M}} \bs p_\stwo ,
  \end{split}
\end{equation}
and $\vrel = \prel \sqrt{T/\mu_{\sone \stwo}}$. Here, $M$ is the total mass and $\mu_{\sone \stwo}$ is the reduced mass. For colliding ions with velocities $\bs v_\sone$ and $\bs v_\stwo$, the normalized velocities are ${\bs p_\sone = \bs v_\sone/\vthone}$ and ${\bs p_\stwo = \bs v_\stwo/\vthtwo}$. 
The function $\varphi_{\sone \stwo}$ is a dimensionless function of $\ptot$ and $\prel$ defined by
\begin{equation}
  \label{eq_varphi_defn}
  \varphi_{\sone \stwo} = \frac{f_\sone(p_\sone) f_\stwo(p_\stwo)}{f_M(p_\sone) f_M(p_\stwo)} . 
\end{equation}

As in \S\ref{sec_background}, the integrand in \eqref{eq_reac_two_spec} is dominated by the Gamow peak located at $\prel = b^{1/3}$. Unlike the single-reactant case, however, the two reactant species have different velocities $p_{\sone,*}$ and $p_{\stwo,*}$ at the Gamow peak, namely
\begin{equation}
  \label{eq_p_star_two_spec}
  p_{\sone,*} = \sqrt{\frac{m_\stwo}{M}} b^{1/3} \quad \text{and} \quad p_{\stwo,*} = \sqrt{\frac{m_\sone}{M}} b^{1/3} .
\end{equation}
%Note that \eqref{eq_p_star_two_spec} implies that $v_{\sone,*}/v_{\stwo,*} = m_\stwo/m_\sone$ (where $v_{\sone,*} = p_{\sone,*} \vthone$ and $v_{\stwo,*} = p_{\stwo,*} \vthtwo$). 
When two ions of different masses fuse, the lighter ion has, on average, a much larger velocity than the heavier ion, even relative to their respective thermal velocities. 
In tandem with the strong velocity dependence of the collision frequency in plasma, this means that kinetic dynamics are more likely to affect the fusion reactivity through perturbations to the light-ion distribution than to the heavy-ion distribution.

In addition to their different velocities at the Gamow peak, the two ion species have different collision frequencies and mean free paths. The difference is particularly stark when the ion species have charge states $Z_\sone \neq Z_\stwo$. 
From \eqref{eq_nu_sone_stwo}, the ratio of the two species' thermal mean free paths is
\begin{equation}
  \label{eq_lambda_ratio}
  \frac{\lambda_{\sone,\mr{th}}}{\lambda_{\stwo,\mr{th}}} = \frac{Z_\stwo^2}{Z_\sone^2} \paren{ \frac{n_\stwo Z_\stwo^2 + n_\sone Z_\sone^2 \sqrt{\frac{2 m_\sone}{m_\sone + m_\stwo}}}{n_\sone Z_\sone^2 + n_\stwo Z_\stwo^2 \sqrt{\frac{2 m_\stwo}{m_\sone + m_\stwo}}} } ,
\end{equation}
meaning that a difference in charge state can lead to a large difference in the Gamow-Knudsen numbers of the two species, even when their masses are comparable. For consistency with the collision model described in \S\ref{sec_background}, the dimensionless collision-frequency coefficient $\widehat \nu_\sone$ becomes 
\begin{equation}
  \label{eq_nu_hat_cie}
  \widehat \nu_\sone(p_\sone) = \frac{1}{1 + p_\sone^3} + \frac{(Z_\sone n_\sone + Z_\stwo n_\stwo)\sqrt{\frac{m_e}{m_\sone}}}{Z_\sone^2 n_\sone + Z_\stwo^2 n_\stwo \sqrt{\frac{2m_\stwo}{m_\sone + m_\stwo}}} . 
\end{equation}

We now proceed with the generalization of \eqref{eq_G_k_asymptotic} to two-species reactions. A straightforward extension of \eqref{eq_Phi_peak} yields
\begin{widetext}
\begin{equation}
  \label{eq_Phi_two_spec}
  \Phi \sim \frac{1}{V} \int d^3 x \left\langle 1 + \frac{f_{\sone,1}(\bs x, p_{\sone,*} \bs{\hat r})f_{\stwo,1}(\bs x, -p_{\stwo,*}\bs{\hat r})}{\fm(p_{\sone,*}) \fm(p_{\stwo,*})} + \frac{f_{\sone,2}(\bs x, p_{\sone,*}\bs{\hat r})}{\fm(p_{\sone,*})} + \frac{f_{\stwo,2}(\bs x, p_{\stwo,*}\bs{\hat r})}{\fm(p_{\stwo,*})} \right\rangle_\Omega   .
\end{equation} 
%\end{widetext}
It is useful to cast \eqref{eq_Phi_two_spec} in terms of the utility function defined in \eqref{eq_Phi_E_G}, generalized to the case of multiple reactant species. We define the average ion mass $\overline m = (n_\sone m_\sone + n_\stwo m_\stwo)/(n_\sone + n_\stwo)$ and the average thermal velocity $\overline v_\mr{th} = \sqrt{T/\overline m}$. Then the reactivity enhancement factor can be written as
\begin{equation}
  \label{eq_Phi_E_G_two_spec}
  \Phi \sim 1 + 2 \int_0^\infty dk E(k) G(k) ,
\end{equation}
where $E(k) = E_\sone(k) \vthone^2/\overline v_\mr{th}^2$. 
Let $\gamma_{\sone,*} = \gamma_\sone(k, p_{\sone,*})$ and $\gamma_{\stwo,*} = \gamma_\stwo(k, p_{\stwo,*})$ be the Gamow-Knudsen numbers for the two species. Using \eqref{eq_f1} and \eqref{eq_f2_avg}, a formula for $G(k)$ follows from \eqref{eq_Phi_two_spec} by a procedure similar to that laid out in Ref.~\citenum{Fetsch_Fisch_2025b}. 
As an intermediate step, note that the term in \eqref{eq_Phi_two_spec} corresponding to collisions between ions from the first-order perturbed distributions of each species depends on the integral
%\begin{widetext}
\begin{equation}
  \label{eq_f1f1_integral}
  \begin{split}
  \left\langle \wt f_{\sone,1}(p_{\sone,*} \bs {\hat r},\bs k) \wt f_{\stwo,1}(-p_{\stwo,*} \bs {\hat r},\bs k) \right\rangle_\Omega \propto \int_{-1}^1 d\xi (1-\xi^2) \frac{ \gamma_{\sone,*} \gamma_{\stwo,*}\xi^2(1 -\gamma_{\sone,*} \gamma_{\stwo,*}\xi^2)}{(1 + \gamma_{\sone,*}^2\xi^2)(1 + \gamma_{\stwo,*}^2\xi^2)} = &-\frac{2(2\gamma_{\sone,*}^2 \gamma_{\stwo,*}^2 + 3\gamma_{\sone,*}\gamma_{\stwo,*} + 3\gamma_{\sone,*}^2 + 3\gamma_{\stwo,*}^2)}{3\gamma_{\sone,*}^2 \gamma_{\stwo,*}^2} \\
  &- \frac{2\gamma_{\stwo,*}(\gamma_{\sone,*}^2+1)\tan^{-1}(\gamma_{\sone,*})}{\gamma_{\sone,*}^3(\gamma_{\sone,*}-\gamma_{\stwo,*})}  \\
  &+ \frac{2\gamma_{\sone,*}(\gamma_{\stwo,*}^2+1)\tan^{-1}(\gamma_{\stwo,*})}{\gamma_{\stwo,*}^3(\gamma_{\sone,*}-\gamma_{\stwo,*})}  .
  \end{split}
\end{equation}
Then, to leading order in $b$, the two-species utility function is
\begin{equation}
  \label{eq_G_k_two_spec}
  \begin{split}
  G(k) =  & \frac{m_\sone m_\stwo}{(m_\sone + m_\stwo) \overline m} \frac{b^{2/3}}{4}\Bigg [ -\frac{2(2\gamma_{\sone,*}^2 \gamma_{\stwo,*}^2 + 3\gamma_{\sone,*}\gamma_{\stwo,*} + 3\gamma_{\sone,*}^2 + 3\gamma_{\stwo,*}^2)}{3\gamma_{\sone,*}^2 \gamma_{\stwo,*}^2} \\
  - & \frac{2\gamma_{\stwo,*}(\gamma_{\sone,*}^2+1)\tan^{-1}(\gamma_{\sone,*}) }{\gamma_{\sone,*}^3(\gamma_{\sone,*}-\gamma_{\stwo,*})} 
  + \frac{2\gamma_{\sone,*}(\gamma_{\stwo,*}^2+1)\tan^{-1}(\gamma_{\stwo,*})}{\gamma_{\stwo,*}^3(\gamma_{\sone,*}-\gamma_{\stwo,*})} 
  \\
  + & \frac{4\gamma_{\sone,*}^3 - 6(\gamma_{\sone,*}^2 + 1)\tan^{-1}(\gamma_{\sone,*}) + 6\gamma_{\sone,*}}{3\gamma_{\sone,*}^3} 
  +\frac{4\gamma_{\stwo,*}^3 - 6(\gamma_{\stwo,*}^2 + 1)\tan^{-1}(\gamma_{\stwo,*}) + 6\gamma_{\stwo,*}}{3\gamma_{\stwo,*}^3} 
  \Bigg ] .
  \end{split}
\end{equation}
\end{widetext}
In the case of a single reactant species, where $\gamma_{\sone,*} \to \gamma_{\stwo,*}$ and $m_\sone \to m_\stwo$, $G(k)$ reduces to the expression found in \eqref{eq_G_k_asymptotic}. The equivalence of \eqref{eq_G_k_two_spec} and \eqref{eq_G_k_asymptotic} in the single-species limit can be verified graphically or analytically. In the latter case, note that the apparent singularity in the terms in the second line of \eqref{eq_G_k_two_spec} vanishes when the two terms are added together; applying L'H\^opital's rule to these terms and adding the other components of \eqref{eq_G_k_two_spec} yields the expected single-species utility function.

\subsection{Results}

\begin{figure}[!hb]
  \centering
  \begin{subfigure}{\columnwidth}
    \centering
    \makebox[\columnwidth][bc]{%
      \hspace*{0pt}%
      \begin{tikzpicture}
        \node[inner sep=0] (imgA) {\includegraphics[width=0.95\columnwidth]{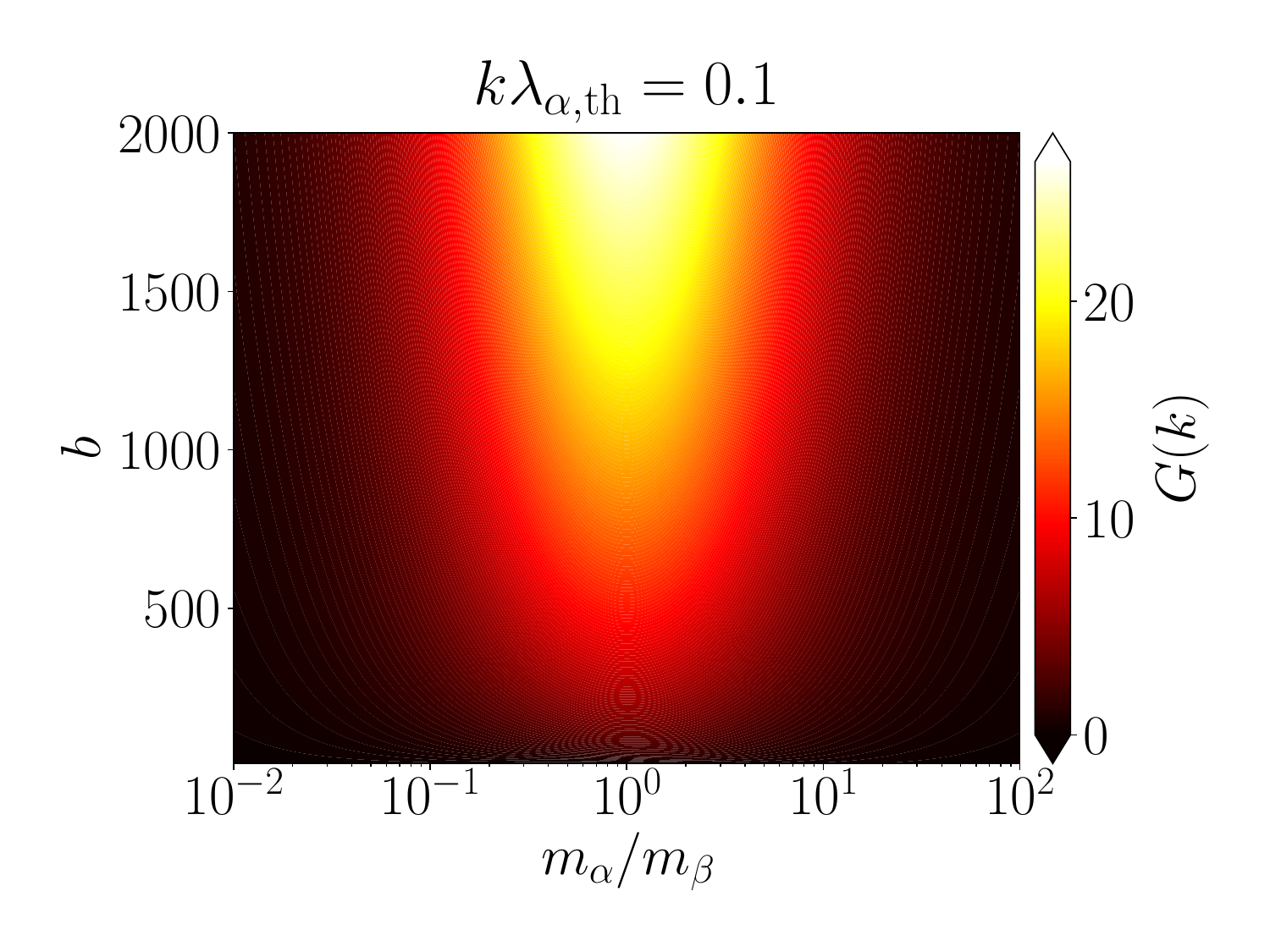}};
        \node[anchor=north west, font=\fontsize{\subfiglabelsizept pt}{\subfiglabelleadingpt pt}\selectfont, white] at ([xshift=0.18\linewidth,yshift=-0.11\linewidth]imgA.north west) {(a)};
      \end{tikzpicture}%
    }
  \end{subfigure}
  \vspace{0em}
  \begin{subfigure}{\columnwidth}
    \centering
    \makebox[\columnwidth][bc]{%
      \hspace*{0pt}%
      \begin{tikzpicture}
        \node[inner sep=0] (imgB) {\includegraphics[width=0.95\columnwidth]{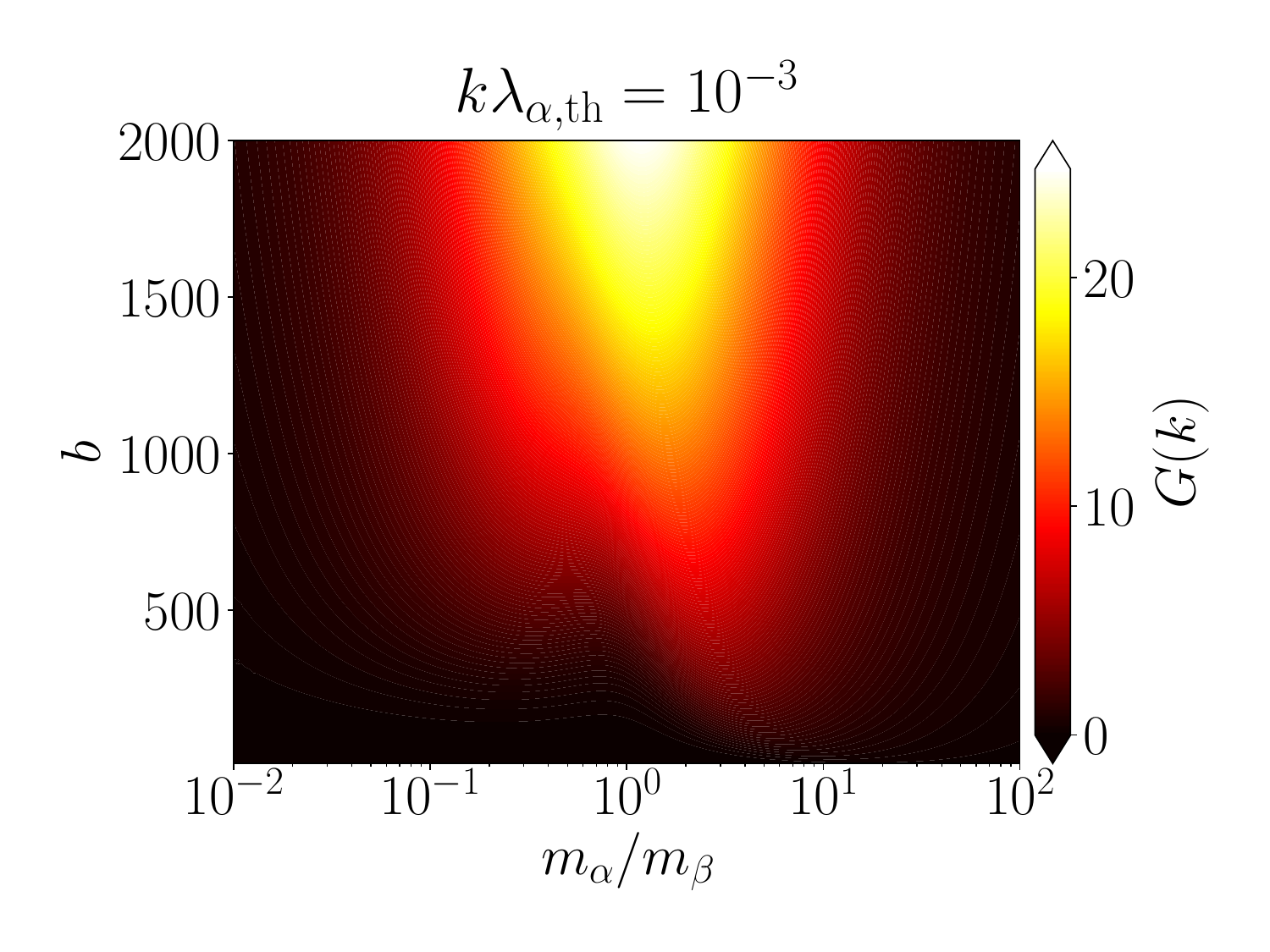}};
        \node[anchor=north west, font=\fontsize{\subfiglabelsizept pt}{\subfiglabelleadingpt pt}\selectfont, white] at ([xshift=0.18\linewidth,yshift=-0.11\linewidth]imgB.north west) {(b)};
      \end{tikzpicture}% 
    }
  \end{subfigure}
  \caption{\justifying Reactivity-enhancement utility function $G(k)$ for reactions between species with mass ratio $m_\sone/m_\stwo$ and Gamow parameter $b$ in flows with $k\lth = 0.01$ (a) and $k\lth = 10^{-3}$ (b). The charge ratio varies as $Z_\sone/Z_\stwo = m_\sone/m_\stwo$. At the $k$ values shown, the result is weakly dependent on the absolute mass and charge of species $\sone$; for illustration, $m_\sone = 40 m_p$ ($m_p$ is the proton mass) and $Z_\sone = 20$ are used here, although the results shown in this figure depend only weakly on these parameters.}
  \label{fig_map_m_b} 
\end{figure}

\begin{figure}[t]
  \centering 
  \includegraphics[width=0.95\columnwidth]{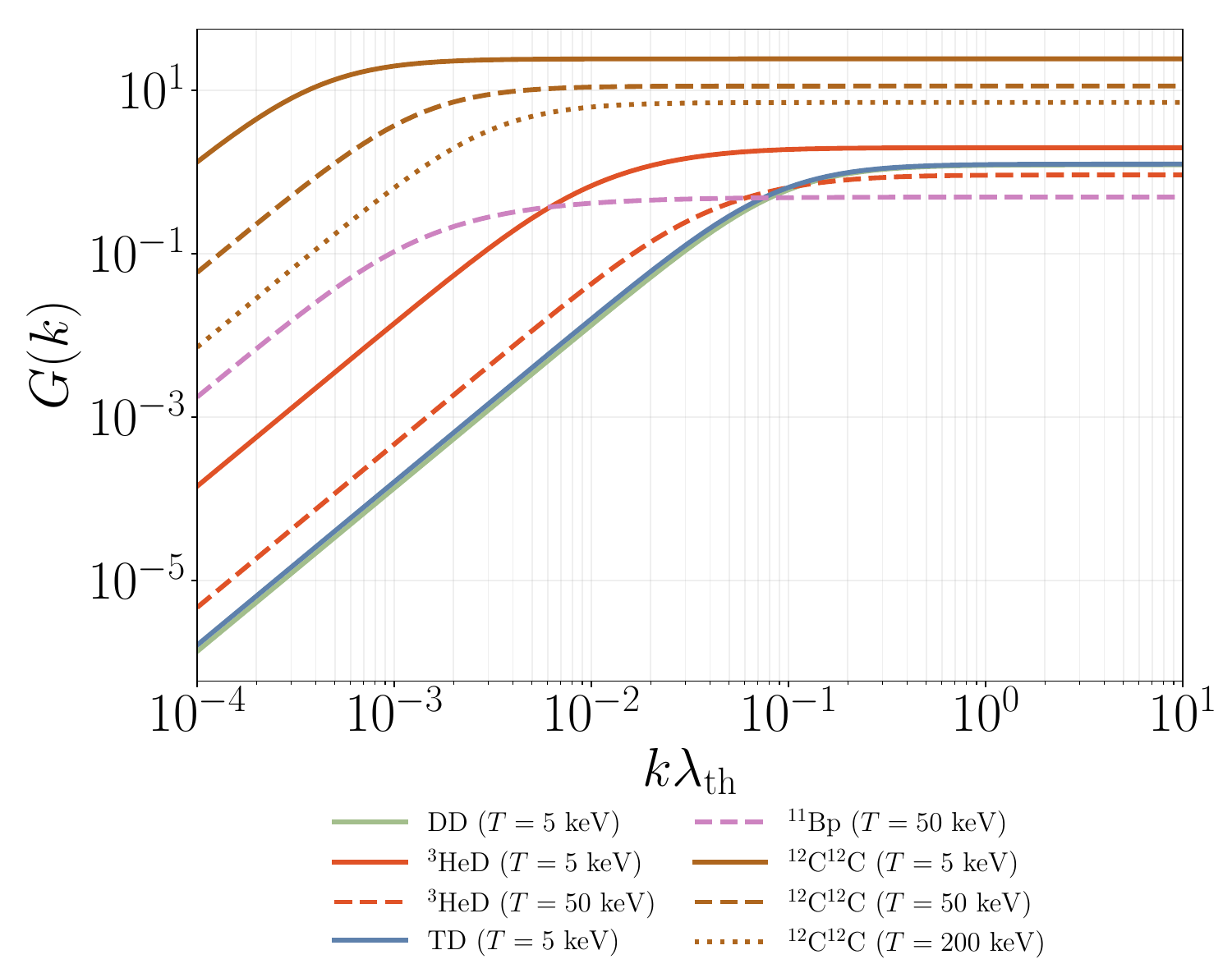}
  \caption{\justifying Reactivity-enhancement utility function $G(k)$ for a variety of reactions of interest in laboratory and astrophysical fusion: DD (deuterium-deuterium), D\textsuperscript{3}He (deuterium-helium-3), DT (deuterium-tritium), p\textsuperscript{11}B (proton-boron-11), and \textsuperscript{12}C\textsuperscript{12}C (carbon-12-carbon-12). For each reaction, species $\sone$ is identified with the heavier reactant, and $k$ is normalized to $\lthsone$.}
  \label{fig_G_k_reactions}
\end{figure}

The dependence of $G(k)$ on the mass ratio and Gamow parameter is shown in Fig.~\ref{fig_map_m_b} for flows on scales such that $k\lth = 0.01$ (a) and $k\lth = 10^{-3}$ (b). The charge ratio is assumed to vary with the mass ratio according to $Z_\sone/Z_\stwo = m_\sone/m_\stwo$. 
For simplicity, the thermal mean free path of species $\sone$ is used as the length scale relative to which the flow wavenumber is measured. In Fig.~\ref{fig_map_m_b}, the mass of species $\sone$ is fixed at forty times the proton mass, and the charge state is fixed at $Z_\sone = 20$ (i.e. species $\sone$ is \textsuperscript{40}Ca). These mass and charge values are chosen simply to illustrate a wide range of mass ratios. At the $k$ values used in Fig.~\ref{fig_map_m_b}, the result depends very weakly on the absolute mass and charge of species $\sone$ since the primary dependence on $m_\sone$ and $Z_\sone$ is already absorbed into the normalization factors; the effect of varying $m_\sone$ and $Z_\sone$ is to shift slightly the velocity at which ion-electron collisions become important relative to ion-ion collisions. At smaller $k$ values, where even very fast particles cross only a small fraction of a flow wavelength between collisions, the shape of $G(k)$ in $(m_\sone/m_\stwo, b)$ space becomes more sensitive to the absolute mass and charge of species $\sone$ because the role of ion-electron collisions in reducing the slope of $\lambda(w)$ becomes more significant in determining the size of the reactivity enhancement. 

As found in Refs.~\citenum{Fetsch_Fisch_2025a,Fetsch_Fisch_2025b}, and anticipated by arguments in \S\ref{sec_sfre}, $G(k)$ is largest when the Gamow parameter is large, meaning that the Gamow peak is further out on the tail. 
In Fig.~\ref{fig_map_m_b}, it is also apparent that $G(k)$ peaks when the reactants have approximately equal mass and falls off quickly as the disparity in mass increases. Interestingly, in the longer-length scale case shown in panel (b), an asymmetry appears in $G(k)$ with respect to the mass ratio. 
The region of increased $G(k)$ in the lower right quadrant of panel (b) ($m_\sone/m_\stwo > 1$ and $b < 1000$) results from light ions of species $\stwo$, which have relatively small charges, being able to travel much further across flow gradients than the heavier ions of species $\sone$. (The asymmetry almost entirely disappears when $Z_\sone/Z_\stwo$ is held constant.) As the mass ratio continues to increase, $G(k)$ again decreases. In both the $m_\sone \gg m_\stwo$ and $m_\sone \ll m_\stwo$ limits, the SFRE disappears because the heavy ion species carries nearly all of the flow energy, while the light ion species is the one whose velocity is more important in determining fusion reactivity; cf. \eqref{eq_p_star_two_spec}. In this limit, the flow velocity becomes negligible relative to the thermal velocity of the light ion species, meaning that the boost in peculiar velocity available by crossing a flow gradient becomes increasingly meager.

In Fig.~\ref{fig_G_k_reactions}, $G(k)$ is shown for several reactions that are of relevance in fusion experiments and in astrophysical systems. In each reaction, the heavier species is identified with $\sone$ and the lighter with $\stwo$. To illustrate the temperature dependence of the SFRE, some reactions are shown at multiple temperatures. Reactants are assumed to be equimolar. For all reactions shown, the asymptotic scalings of $G(k)$ at small and large $k$ are straight lines in log-log space with slopes $2$ and $0$ respectively. 
The DD and DT curves remain very close to each other at all $k$, but the ratio of the utility function for D\textsuperscript{3}He to that for DD (or DT) varies significantly with $k$. This dependence may have utility as a diagnostic in ICF experiments, where Fig.~\ref{fig_G_k_reactions} suggests that, in principle, yield ratios could be used to infer information about the length scales of turbulent flows.

As a concrete application of \eqref{eq_G_k_two_spec}, we evaluate $G(k)$ under some conditions of relevance to ICF experiments. The following parameters are chosen only for the sake of illustration and do not represent optimized conditions for a large reactivity enhancement (cf. Ref.~\citenum{Fetsch_Fisch_2026}). Consider a DT plasma with ${\rho = 100~\mr{g/cm}^3}$ and ${T = 3~\mr{keV}}$, which may be found in a sub-ignition indirect-drive experiment or in some regions of an imploding target prior to ignition. Suppose that flows are driven in the plasma with a length scale ${L = 2\pi/k = 5~\mu\mr{m}}$. While this could, in principle, be a periodic, structured flow, it can more realistically be conceived of as a turbulent or quasi-turbulent flow dominated by eddies with a characteristic size around $5~\mu\mr{m}$. Let the energy of the flow be ${10\%}$ of the thermal energy, meaning that ${E = 0.3}$. While the Knudsen number of this flow is modest at ${\mr{Kn} \approx 0.004}$, the Gamow-Knudsen numbers\footnote{The definitions in \eqref{eq_Kn_sfre} and \eqref{eq_gamma_defn} exaggerate the difference by a factor of $2\pi$ because $\gamma = 2\pi \mr{Gk}$, but, in any event, the thermal and Gamow scales are separated by well over an order of magnitude in this example.} for both species are appreciable: $\gamma_{D,*} \approx 0.7$ and $\gamma_{T,*} \approx 0.4$. Under these conditions, ${G(k) \approx 0.22}$, and the reactivity enhancement factor is ${\Phi \approx 1.13}$. The viscous dissipation time is $\tau_\eta \approx 50~\mr{ps}$, which is comparable to a typical burn time (although an igniting target would subsequently heat up and develop a higher viscosity, leading to much faster dissipation of turbulent energy later in the burn).  
As an alternative scenario, consider instead a DT plasma akin to that in the fast-ignition scenario described in Ref.~\citenum{Fetsch_Fisch_2025a}. Let $\rho = 300~\mr{g/cm}^3$ and $T = 8~\mr{keV}$. Suppose that flows are driven with a length scale of $L = 2~\mu\mr{m}$ and an energy of $25\%$ of the thermal energy. In this system, $G(k) \approx 0.58$ and $\Phi \approx 1.87$. The dissipation time is $\tau_\eta \approx 3~\mr{ps}$, meaning that the flows would not be expected to survive the burn duration, but the reactivity enhancement may be present for a long enough time to ``jump-start'' the burn, particularly in a fast-ignition scheme\cite{Fetsch_Fisch_2025a}. 
The subsequent rapid dissipation of the flow heats the plasma -- and, notably, transfers energy first to the ions before they cool on the electrons -- which can have further favorable consequences for rapidly igniting a target \cite{Davidovits_Fisch_2016a,Davidovits_Fisch_2018}.

\section{Viscous dissipation}
\label{sec_visc_diss}

The enhancement of fusion reactivity by sheared flows is inherently inseparable from the viscous dissipation of those flows. In fact, the mechanism described in \S\ref{sec_sfre} underlying the SFRE is precisely the mechanism that gives rise to viscosity in weakly coupled plasma. However, the separation of scales between the thermal bulk and the Gamow window means that these phenomena can be separated in magnitude by an asymptotically large factor. 

To illustrate this point, consider a plasma consisting of a single ion species $\sone$. 
Let the characteristic time for reactants to fuse be $\tau_f$, where
\begin{equation}
  \label{eq_tau_f_defn}
  \tau_f = \frac{1}{n\avg{\sigma v}} ,
\end{equation}
and let $\tau_\sone$ be the ion-ion collision time for species $\sone$. 
To quantify the relative rates of Coulomb collisions and fusion reactions, we define a parameter $\epsilon$ such that
\begin{equation}
  \label{eq_eps_defn}
  \epsilon = \frac{\tau_\sone}{\tau_f} .
\end{equation}
\begin{comment}
which can be written, for reactions that are adequately approximated by \eqref{eq_reac_maxwellian_approx}, as
  \begin{equation}
  \label{eq_eps_explicit}
  \epsilon \sim \frac{3}{4\sqrt{\pi}} \frac{\sqrt{m_i} T^{3/2}}{Z^4 e^4 \ln \Lambda} \sqrt{\frac{3}{2}} \frac{\vth}{A b^{1/3}} \frac{1}{S(\vth b^{1/3}/\sqrt{2})} e^{\frac{3}{2}b^{2/3}} ,
\end{equation}
\begin{equation}
  \label{eq_E_G_defn_TEMPORARY}
  E_\mr{G} = 2\pi^2 \frac{\mu Z_1^2Z_2^2 e^4}{\hbar^2}
\end{equation}
\begin{equation}
  \label{eq_eps_explicit}
  \epsilon \sim \frac{3^{3/2}}{2^{5/2}\sqrt{\pi}} \frac{b^2(2E_G)^2}{Z^4e^4\ln\Lambda b^{1/3} A} \frac{ e^{\frac{3}{2}b^{2/3}}}{S(\vth b^{1/3}/\sqrt{2})} ,
\end{equation}
\begin{equation}
  \label{eq_eps_explicit}
  \epsilon \sim \frac{(6\pi)^{3/2} \mu_{\sone \sone}^2 Z^4e^4 }{\hbar^2 \ln \Lambda} \frac{b^{5/2}}{A} \frac{ e^{\frac{3}{2}b^{2/3}}}{S(\vth b^{1/3}/\sqrt{2})} ,
\end{equation}
\end{comment}
For most laboratory and astrophysical fusion plasmas, $\epsilon \ll 1$, as discussed in \S\ref{sec_background}. 
The number of ion-ion collisions during the characteristic time required for fusion reactions to occur is of order $1/\epsilon \gg 1$.
Hence, even if a fusion plasma is initiated with a spatially uniform kinetic perturbation that enhances its reactivity, the plasma will relax to a Maxwellian before the transient enhancement factor can translate into a large increase in the absolute number of reactions. If $\Phi \sim \mc O(1)$ and the confinement time $\tau_c$ is of order $\tau_f$, then the increase in fusion yield $\Delta Y$, normalized to the yield $Y_0$ produced by Maxwellian reactants, scales as 
\begin{equation}
  \label{eq_Delta_Y_local}
  \frac{\Delta Y}{Y_0} \sim \frac{(\Phi - 1)}{\widehat{\nu}_*} \epsilon
\end{equation}
%$\Delta Y/Y_0 \sim \mc O(\epsilon)$.
where the factor of $\widehat{\nu}_* = \widehat{\nu}(p_*)$ accounts for the fact that the portion of the perturbation at the Gamow peak is less collisional and relaxes on a time scale longer than $\tau_\sone$, provided that kinetic instabilities do not intervene.

To achieve a significant increase in yield through kinetic enhancements, the plasma must be continuously driven out of equilibrium. The SFRE allows this driving to be done by inhomogeneous flows. 
In a sheared flow, the counterstreaming populations of thermal reactants are separated in space, meaning that the flow does not relax on the collisional time scale $\tau_\sone$. Instead, relaxation is governed by the diffusive transport of thermal ions across the flow gradient, which occurs on a much longer viscous time scale $\tau_\eta$. 
Notably, fast ions cross the flow gradient more quickly than thermal ions do and have longer mean free paths, leading to a larger fractional enhancement in the tail of the distribution than in the bulk. Serendipitously, then, the perturbation can be large in the region where fusion reactivity is determined while being small in the region where viscosity is determined and where most dissipation takes place.

Consider a plasma with a flow satisfying the conditions described in \S\ref{sec_turb_reac} (incompressible, subsonic, and with small Knudsen number), and for simplicity let the flow be described by a single wavenumber $k$. The viscous dissipation time is
\begin{equation}
  \label{eq_tau_eta}
  \tau_\eta = \frac{1}{2 k^2 \eta} ,
\end{equation}
where, as in \S\ref{sec_turb_reac}, $\eta = c_\eta \lth^2 \nu_\sone$ is the kinematic viscosity.

If the flow produces a reactivity enhancement factor $\Phi$, and the plasma is confined for a time $\tau_c \sim \tau_f$, then the fractional increase in yield can be estimated as ${\Delta Y/Y_0 \sim (\Phi - 1) \tau_\eta/\tau_f}$. This estimate neglects any positive feedback resulting from the increased fusion power and may therefore underestimate the yield increase in ignited systems, where a small boost in self-heating during early stages of the burn can be exponentially amplified at later stages. 
As an estimate of the importance of the SFRE in a given system, let ${\Psi(k) = \Delta Y / E(k) Y_0}$ be the increase in yield per unit of flow energy in modes of wavenumber $k$. Then, using \eqref{eq_Phi_E_G_two_spec} and \eqref{eq_tau_eta}, we have
\begin{equation}
  \label{eq_Psi_G}
  \Psi(k) \sim \frac{G(k)}{c_\eta k^2\lth^2} \epsilon .
\end{equation}
Considering a single species for simplicity, \eqref{eq_Psi_G} can be reduced to an elegant form in the case of a small Gamow-Knudsen number, where $\gamma_* \ll 1$. Expanding \eqref{eq_G_k_asymptotic} in this limit, we find $G(k) \sim b^{2/3} \gamma_*^2 / 10$. Noting further that ${c_\eta \approx 0.96}$ for a single ion species in unmagnetized plasma, \eqref{eq_Psi_G} reduces to
\begin{equation}
  \label{eq_Psi_lstar}
  \Psi(k) \sim \frac{b^{2/3}}{10} \paren{\frac{\lambda_*}{\lth}}^2 \epsilon .
\end{equation}
The quantity $\lambda_*/\lth$, or equivalently the ratio of the Gamow mean free path to the thermal mean free path, is typically a large parameter, with a $b$-dependence ranging from $b^{1/3}$ to $b^{4/3}$ depending on the electron density and electron-ion mass ratio, as seen in Fig.~\ref{fig_lambda_star_ratio}. Hence, \eqref{eq_Psi_lstar} leads to scalings between $\Psi \propto b^{4/3} \epsilon$ and $\Psi \propto b^{10/3} \epsilon$. 
Because $b^{4/3}$ is a large parameter -- and $b^{10/3}$ is an even larger one -- $\Psi$ can be of order unity or greater, even when $\epsilon \ll 1$. 
%This illustrates the key distinction between the SFRE and many other kinetic reactivity enhancements. In sheared flows, the counterstreaming thermal populations are kept separated in space.

The significance of the fact that the reactivity enhancement decays on the viscous time scale, rather than on the collisional time scale -- i.e. that the dissipation is a result of a slow diffusive process as opposed to a fast process of local equilibration -- is underscored by comparing $\Psi(k)$ computed in a sheared flow to the analogous quantity $\Psi'$ computed in a spatially uniform non-Maxwellian plasma. Let $\Psi' = \Delta Y' / E' Y_0$, where $E'$ is the energy of the perturbation, normalized to the temperature, and $\Delta Y'$ is the increase in yield produced by the perturbation. Letting $\Phi'$ be the reactivity enhancement factor produced by the perturbation and using \eqref{eq_Delta_Y_local} gives ${\Psi' \sim (\Phi' - 1) \epsilon/\widehat{\nu}_* E'}$. Then, using \eqref{eq_Psi_lstar} and recalling that $\lambda_* = b^{1/3} \lth /\sqrt{2} \widehat{\nu}_*$, the ratio of the normalized increases in yield is
\begin{equation}
  \label{eq_Psi_ratio}
  \frac{\Psi'}{\Psi(k)} \sim \frac{14}{b} \paren{\frac{\lth}{\lambda_*}} \frac{\Phi'-1}{E'} .
\end{equation}
It follows from \eqref{eq_Psi_ratio} that the SFRE outperforms local non-Maxwellian perturbations, in the sense of producing a larger time-integrated yield increase for the same amount of energy, under the condition 
\begin{equation}
  \label{eq_Phi_E_Psi_Psiprime_condition}
    \frac{\Phi' - 1}{E'} \lesssim \frac{b}{14} \paren{\frac{\lambda_*}{\lth}} .
\end{equation}
The left-hand and right-hand sides of \eqref{eq_Phi_E_Psi_Psiprime_condition} can be seen in Fig.~\ref{fig_M_j_norm} and Fig.~\ref{fig_lambda_star_ratio}, respectively. Noting the convenient mnemonic that $b \approx 30$ at $T = 3~\mr{keV}$ in DD or DT plasma, comparison of Figs.~\ref{fig_M_j_norm} and \ref{fig_lambda_star_ratio} indicates that the right-hand side of \eqref{eq_Phi_E_Psi_Psiprime_condition} is typically larger in ICF devices. 
This means, surprisingly, that the time-integrated yield increase produced by driving a sheared flow and then allowing it to decay is larger than that produced by expending the same amount of energy to increase the number of particles around the Gamow peak and then allowing the plasma to relax toward equilibrium. 
More generally, using \eqref{eq_Phi_example_beam} and \eqref{eq_M2_example} lets us reduce \eqref{eq_Phi_E_Psi_Psiprime_condition} to the form ${6^{3/2} b^{-4/3} \exp(b^{2/3}/4) \lesssim b \lambda_*/14\lth}$. It follows that ${\Psi(k) > \Psi'}$ when
\begin{equation}
  \label{eq_b_conditions_Psi}
  1 \ll b \lesssim 524 \quad \text{or} \quad 1 \ll b \lesssim 431
\end{equation}
representing, respectively, cases where the Gamow mean free path scales as $\lambda_*/\lth \sim b^{4/3}/4$ and as ${\lambda_*/\lth \sim 30 b^{1/3}}$, depending on the significance of ion-electron collisions at the Gamow peak (cf. \eqref{eq_nu_p}). Within the range given in \eqref{eq_b_conditions_Psi}, the SFRE increases yield more efficiently than simply driving an enhanced tail at the Gamow peak. Below this range, the asymptotic large-$b$ theory breaks down; above this range, perturbations at the Gamow peak increase yield more efficiently than sheared flows. 

Note that, even within the wide Gamow-parameter range described by \eqref{eq_b_conditions_Psi}, this result does not categorically preclude general kinetic perturbations from outperforming the SFRE. For instance, driving a beam-like distribution at a velocity higher than $v_*$ would enhance reactivity more efficiently than the perturbations considered here and in \S\ref{sec_background}, which are centered on the Gamow peak. For such beams, the Gamow-peak approximation breaks down because the perturbation no longer varies slowly relative to the exponential parts of the reactivity integrand. Distributions of this type may be susceptible to kinetic instabilities\cite{VanZeeland_et_2021}, reducing the time scale on which they enhance reactivity to a value well below the collisional equilibration time. 
Moreover, this analysis makes no accounting for the difficulty of generating the perturbations considered here. In fact, it may be seen as an advantage of the SFRE that its impact can be optimized simply by stirring the plasma in certain ways, in contrast to the finer control (``phase-space engineering'') required for the optimization of some other kinetic effects \cite{Fisch_1987,Kolmes_Fisch_2024,Qin_2024,Graves_Chapman_Coda_Lennholm_Albergante_Jucker_2012,Updike_Bohlsen_Qin_Fisch_2025,Qin_Kolmes_Updike_Bohlsen_Fisch_2025}.

Setting these details aside, the result in \eqref{eq_b_conditions_Psi} is striking. Intuition would suggest that, if the tools are available to exert fine-grained control of the velocity-space distribution of a fusion plasma, then exercising this control to place more ions around the Gamow peak is likely to be a more efficient use of energy than the comparatively blunt instrument of stirring hydrodynamic motion in the plasma. 
But remarkably, when $b$ satisfies the condition \eqref{eq_b_conditions_Psi}, the intuitive picture is not borne out. In this range, the much longer persistence of sheared flows compared to spatially uniform non-equilibrium distributions outweighs the benefit of the latter concentrating the perturbation energy in the tail of the ion distribution. 
Physically, the key difference is the following: when a perturbed distribution is set up and allowed to evolve, there is nothing to replenish the non-equilibrium population in the tail -- as soon as the tail ions equilibrate with the bulk, they remain thermal indefinitely. In a sheared flow, only a fraction of tail ions have the right velocity to cross the flow gradient, and this population is attenuated by collisions so that the fraction of ions reaching regions with substantially different background flow is further reduced. Crucially, however, this tail population is being continually replenished by velocity-space diffusion of bulk ions at each point in space. This replenishment continues until the flow has dissipated, which happens orders of magnitude more slowly than collisions do. Thus, while in a sheared flow a smaller number of ions is available to enhance the reactivity at each moment in time, the time-integrated effect on the reactivity can be significantly larger. 
Note that this curious behavior is only relevant if the reactivity enhancement by the SFRE, or by a conventional kinetic perturbation, is substantially larger than the increase in thermal reactivity from the heating that occurs when the perturbation dissipates and delivers its energy to the plasma, or if the plasma is in contact with a thermostat of some kind that prevents the ions from heating up as the perturbation dissipates.

In some cases, it is of practical interest to quantify the viscous lifetime of flows as a function of the reactivity enhancement that they produce. For instance, in an experiment designed to measure the enhancement of reactivity by turbulence -- or to exploit that enhancement to achieve high gain in an ICF implosion -- it is important to know whether the flows generating a given enhancement factor persist on time scales long enough to produce a measurable effect. 
Let $L_e(\Phi)$ be the length scale such that flows with normalized energy $E$ and wavenumber $k = 2\pi/L_e$ produce an enhancement factor $\Phi$. In the limit where $\Phi - 1 \ll 1$, we know that $L_e$ will be large and can therefore use the small-$\gamma_{\sone,*}$ expansion of $G(k)$ to find
\begin{equation}
  \label{eq_L_e_Phi}
  L_e(\Phi) \sim 2\pi \lambda_{\sone,*} b^{1/3} \sqrt{\frac{E}{5(\Phi - 1)}} .
\end{equation}
Let $\tau_e(\Phi)$ be the viscous lifetime of flows on length scale $L_e(\Phi)$, and suppose that we are in the limit where ${\lambda_{\sone,*}/\lthsone \sim b^{4/3}/4}$.  Then, using \eqref{eq_tau_eta}, we have
\begin{equation}
  \label{eq_tau_eta_Phi}
  \tau_e(\Phi) \sim \tau_\sone \frac{E}{\Phi - 1} \frac{b^{10/3}}{160} .
\end{equation}
From \eqref{eq_tau_eta_Phi}, we see that $\tau_e(\Phi) \propto T^{-1/6}n^{-1}$; the flows necessary to produce a given reactivity enhancement persist for longer times at lower temperatures and lower densities. 
This scaling is not necessarily obvious; based on the scaling of the kinematic viscosity $\eta \propto T^{5/2}n^{-1}$, one might expect that cooler, denser, and more collisional plasmas would be more conducive to the SFRE because they can support fine-scale flows for longer times. In actuality, the shorter Gamow mean free paths in dense plasmas reduce the size of the enhancement; to achieve the same $\Phi$, flows must then be on finer scales. Thus, per \eqref{eq_tau_eta_Phi}, the flows producing a given enhancement factor survive a shorter time in these plasmas. However, in the case where collisions in the Gamow window are dominated by ions, the rapid increase in scale separation between the Gamow mean free path and the thermal mean free path with decreasing temperature means that $\tau_e(\Phi)$ is larger in cooler plasmas.

\section{Discussion}
\label{sec_disc}

The shear flow reactivity enhancement is a striking consequence of the large separation between thermal scales and reactivity-relevant scales in fusion plasmas. 
Distinctively in plasmas, as opposed to in neutral gases or other fluids, the strong velocity dependence of the collision frequency allows fast ions to travel much longer distances than their thermal counterparts. Therefore, even in systems where the Knudsen number is small and kinetic effects may be considered negligible, the population of ions near the Gamow peak may exhibit significant non-thermal features.

The SFRE reproduces aspects of enhanced reactivity driven by counterpropagating ion beams, a possibility that formed the basis for some optimized tokamak designs \cite{Furth_Jassby_1974}. However, obtaining a reactivity enhancement from counterpropagating ion beams requires that the ion populations with large relative velocities be colocated in space, leaving such configurations vulnerable to a variety of kinetic instabilities and to rapid collisional relaxation. By contrast, the SFRE relies on sheared flows in the near-hydrodynamic regime, so that the relatively moving components of the plasma are separated from each other, rather than interpenetrating. The result is that the relative motion might lead to dissipation by viscosity, but the virulent collisionless two-stream instabilities are avoided. The dissipation, meanwhile, is a diffusive process operating over the length scale of the flow, rather than a purely collisional process happening at each point in space, and is therefore much slower than collisional relaxation of interpenetrating beams, even in the absence of kinetic instabilities.

To emphasize the fundamental nature of the SFRE, this work focused on the limit of large Gamow parameter (${b \gg 1}$), where the scale separation underlying the effect becomes asymptotically large. This limit allowed us to approximate the reactivity in \S\ref{sec_background} and facilitated the derivation of a simple analytical formula for the reactivity-enhancement utility function $G(k)$ in \S\ref{sec_turb_reac}. Moreover, we showed in \S\ref{sec_background} that, when $b \gg 1$, perturbations to the ion distribution near the Gamow peak have an asymptotically small effect on fluid moments. For this reason, as we showed in \S\ref{sec_visc_diss}, the time scale of viscous dissipation becomes asymptotically long relative to the time scale of fusion reactions, allowing for significant reactivity enhancements before the flows generating those enhancements are dissipated.

In some astrophysical systems, and in some ICF experiments involving rare reactions between high-$Z$ elements \cite{Adrian_et_2025,Casey_et_2017,Casey_et._2023,Jeet_et_2023}, the Gamow parameter can be in the range of several hundreds or thousands, justifying the use of the large-$b$ limit in this work. 
Under conditions characterizing present high-yield ICF experiments -- for example, ignited experiments at the National Ignition Facility\cite{ICF_collab_2024,Kritcher_et_2024} -- the Gamow parameter is typically more modest ($b \approx 15-30$). The lack of a very large parameter in these cases complicates the theoretical description of the SFRE.  In Ref.~\citenum{Fetsch_Fisch_2025b}, contributions from higher orders in $1/b$ are taken into account to derive a ``corrected utility function,'' which agrees well with reduced continuum-kinetic simulations \cite{Fetsch_Fisch_2025b}. 
The effect of the reactivity enhancement on ICF ignition was studied in Ref.~\citenum{Fetsch_Fisch_2026}, where the Lawson criterion was generalized to a ''turbulent ignition criterion'' applying to hot spots containing fine-scale turbulence. A parameter regime was identified in which an ICF hot spot would ordinarily fail to ignite but could, in fact, ignite if a fraction of its thermal energy were replaced with turbulent energy on appropriately chosen length scales, provided that most turbulent energy could be prevented from ending up in deleterious large-scale eddies \cite{Fetsch_Fisch_2026}.

Designing experiments to reach higher gain by driving turbulent flows on optimal scales would require techniques for the control of ICF turbulence that have not yet been developed, although there is some numerical evidence that such control is attainable\cite{Albright_et_2022,Haines_et_2023,Murphy_et_2021,Davidovits_et_2022,Fetsch_Fisch_2025a,Li_Davidovits_2024}. Particularly large benefits can be expected in fast-ignition or shock-ignition designs, where rapid ion heating is energetically expensive\cite{Fetsch_Fisch_2023}, and where ignition at lower electron temperature leads to a shorter alpha-particle stopping distance, allowing a smaller hot spot\cite{Fetsch_Fisch_2025a}. 
Such a design would likely benefit from synergy with the ``sudden viscous dissipation'' effect \cite{Davidovits_Fisch_2016a,Davidovits_Fisch_2016b,Campos_Morgan_2019,Davidovits_Fisch_2019b}. 

Abstracting away the details of a particular ignition design, the theoretical interest of the SFRE can be put as follows. In a truly hydrodynamic plasma, the question \textit{How should internal energy be partitioned between thermal and turbulent degrees of freedom to maximize the fusion rate?} has a trivial answer. While reactivity can be increased by driving a separation between the ion temperature $T_i$ and the electron temperature $T_e$ so that $T_i > T_e$, such a temperature separation relaxes on the relatively short time scale of ion-electron collisions\cite{Fetsch_Foster_Fisch_2023,Foster_Fetsch_Fisch_2023}. 
As a consequence of the SFRE, the answer to this question becomes complicated. Energy in flows on scales comparable to the Gamow mean free path, as well as the ion temperature, plays a role in determining reactivity. In some cases, then, the optimal energy partition includes a nontrivial turbulent energy spectrum, adding a new degree of freedom to the optimization of fusion-experiment configurations.

\begin{acknowledgments}
This work was supported by the Center for Magnetic Acceleration, Compression, and Heating (MACH), part of the U.S. DOE-NNSA Stewardship Science Academic Alliances Program under Cooperative Agreement DE-NA0004148 and by the National Science Foundation under Grant No. PHY-2308829.

\end{acknowledgments}

\bibliography{sfre_dpp}

\end{document}

%% file: macros.tex
\newcommand{\bs}[1]{\boldsymbol{#1}}

\newcommand{\mc}[1]{\mathcal{#1}}

\newcommand{\half}[0]{\frac{1}{2}}

\newcommand{\paren}[1]{\left(#1\right)}

\newcommand{\wt}{\widetilde}

\newcommand{\avg}[1]{\langle #1 \rangle}

\newcommand{\mr}[1]{\mathrm{#1}}

\newcommand{\vth}{v_\mr{th}}

\newcommand{\lth}{\lambda_\mr{th}}

\newcommand{\sone}{a}
\newcommand{\stwo}{b}

%% file: sfre_dpp.bib
@PREAMBLE{
 "\providecommand{\noopsort}[1]{}" 
 # "\providecommand{\singleletter}[1]{#1}%" 
}

@article{Fetsch_Fisch_2025a, title={Enhancement to Fusion Reactivity in Sheared Flows}, volume={135}, DOI={10.1103/5nll-y8rx}, abstractNote={Sheared flow increases the reactivity of fusion plasma. In unmagnetized plasma with flow gradients comparable to the mean free path of reacting ions, fusion reactivity can be more than doubled. The effect is of particular relevance to inertial confinement fusion (ICF), where it allows implosion kinetic energy to contribute to the fusion burn even before thermalizing. Because colder fuel stops alpha particles more quickly, ignition is possible in a smaller volume, substantially reducing energy requirements in fast-ignition designs.}, number={15}, journal={Physical Review Letters}, publisher={American Physical Society}, author={Fetsch, Henry and Fisch, Nathaniel J.}, year={2025}, month=oct, pages={155101} }

@article{Fetsch_Fisch_2025b, title={Analytical models for the enhancement of fusion reactivity by turbulence}, volume={32}, ISSN={1070-664X}, DOI={10.1063/5.0285620}, abstractNote={The reactivity of fusion plasma depends not only on its local density and temperature but also, through a recently identified kinetic effect, on the relative velocities of nearby fluid elements. Turbulence on fine spatial scales, therefore, enhances fusion reactivity. The enhancement is quantified here for general subsonic turbulent flows. Leveraging this effect in the design of inertial confinement fusion experiments could enable substantial energy savings.}, number={11}, journal={Physics of Plasmas}, author={Fetsch, Henry and Fisch, Nathaniel J.}, year={2025}, month=nov, pages={112703} }

@article{Fetsch_Fisch_2026, title={An ignition criterion for inertial fusion boosted by microturbulence}, volume={33}, ISSN={1070-664X}, DOI={10.1063/5.0295335}, abstractNote={Turbulence on fine spatial scales enhances fusion reactivity, enabling ignition at lower temperature. A modified Lawson-like ignition criterion is derived for inertially confined plasmas harboring turbulent kinetic energy. For some turbulent energy spectra, hot spots ignite at lower energy density and smaller volume. While detrimental mixing effects typically accompany turbulence and obscure these advantages, targets might be engineered to drive flow in regions where it is beneficial. The optimal length scale for this driving is identified, typically lying in the micrometer range.}, number={2}, journal={Physics of Plasmas}, author={Fetsch, Henry and Fisch, Nathaniel J.}, year={2026}, month=feb, pages={020703} }

@article{Fetsch_Fisch_2023, title={Improved ion heating in fast ignition by pulse shaping}, volume={108}, DOI={10.1103/PhysRevE.108.045206}, abstractNote={The fast ignition paradigm for inertial fusion offers increased gain and tolerance of asymmetry by compressing fuel at low entropy and then quickly igniting a small region. Because this hot spot rapidly disassembles, the ions must be heated to ignition temperature as quickly as possible, but most ignitor designs directly heat electrons. A constant-power ignitor pulse, which is generally assumed, is suboptimal for coupling energy from electrons to ions. Using a simple model of a hot spot in isochoric plasma, a pulse shape to maximize ion heating is presented in analytical form. Bounds are derived on the maximum ion temperature attainable by electron heating only. Moreover, arranging for faster ion heating allows a smaller hot spot, improving fusion gain. Under representative conditions, the optimized pulse can reduce ignition energy by over 20%.}, number={4}, journal={Physical Review E}, publisher={American Physical Society}, author={Fetsch, Henry and Fisch, Nathaniel J.}, year={2023}, month=oct, pages={045206} }

@article{Frank_Wright_Rodriguez-Fernandez_Howard_Bonoli_2024, title={Simulating energetic ions and enhanced fusion rates from ion-cyclotron resonance heating with a full-wave/Fokker–Planck model}, volume={31}, ISSN={1070-664X}, DOI={10.1063/5.0204671}, abstractNote={Reproducing fast-ion enhanced fusion rates from ion-cyclotron resonance heating (ICRH) in tokamaks requires the self-consistent coupling of a full-wave solver and a Fokker–Planck solver, which evolves multiple simultaneously resonant ion species. We introduce a new self-consistent model that iterates the TORIC full-wave solver with the CQL3D Fokker–Planck solver using the integrated plasma simulator (IPS). This model evolves the bounce-averaged ion distribution functions in both parallel and perpendicular velocity-space with a quasilinear radio frequency (RF) diffusion operator valid in the ion finite Larmor radius (FLR) limit and the RF electric fields with the resultant non-Maxwellian FLR dielectric tensor. This produces non-Maxwellian ICRH simulations that are fully self-consistent, fast, and interoperable with integrated modeling frameworks, such as TRANSP/GACODE/IPS-FASTRAN. We demonstrate our model’s capabilities by validating it against experimental data in Alcator C-Mod. We then perform the first RF heating simulations of SPARC using self-consistent non-Maxwellian ion distributions to investigate the potential to enhance fusion rates using ion cyclotron resonance heating generated fast ions.}, number={6}, journal={Physics of Plasmas}, author={Frank, S. J. and Wright, J. C. and Rodriguez-Fernandez, P. and Howard, N. T. and Bonoli, P. T.}, year={2024}, month=june, pages={062503} }

@article{Harvey_McCoy_Kerbel_Chiu_1986, title={ICRF fusion reactivity enhancement in tokamaks}, volume={26}, ISSN={0029-5515}, DOI={10.1088/0029-5515/26/1/004}, abstractNote={A systematic study of ICRF fusion reactivity enhancement has been conducted, using a new bounce-averaged two-dimensional Fokker-Planck code. Second-harmonic heating of deuterium in a 50–50 DT plasma is assumed, and the results are obtained as a function of background plasma density and temperature. An enhancement factor of ten is achieved at low Q (= fusion power/RF power), which is important for ion-tail diagnostics, but at Q = 0.5 the enhancement is ⪅ 2. Significant poloidal variations in ion density (up to 14%) and in fusion reactivity (by a factor up to 2.5) are found.}, number={1}, journal={Nuclear Fusion}, author={Harvey, R.W. and McCoy, M.G. and Kerbel, G.D. and Chiu, S.C.}, year={1986}, month=jan, pages={43}, language={en} }

@article{Kolmes_Mlodik_Fisch_2021, title={Fusion yield of plasma with velocity-space anisotropy at constant energy}, volume={28}, ISSN={1070-664X}, DOI={10.1063/5.0050293}, abstractNote={Velocity-space anisotropy can significantly modify fusion reactivity. The nature and magnitude of this modification depends on the plasma temperature, as well as the details of how the anisotropy is introduced. For plasmas that are sufficiently cold compared to the peak of the fusion cross section, anisotropic distributions tend to have higher yields than isotropic distributions with the same thermal energy. At higher temperatures, it is instead isotropic distributions that have the highest yields. However, the details of this behavior depend on exactly how the distribution differs from an isotropic Maxwellian. This paper describes the effects of anisotropy on fusion yield for the class of anisotropic distribution functions with the same energy distribution as a 3D isotropic Maxwellian and compares those results with the yields from bi-Maxwellian distributions. In many cases, especially for plasmas somewhat below reactor-regime temperatures, the effects of anisotropy can be substantial.}, number={5}, journal={Physics of Plasmas}, author={Kolmes, E. J. and Mlodik, M. E. and Fisch, N. J.}, year={2021}, month=may, pages={052107} }

@article{Ye_Zhang_Wan_2025, title={Enhancement of the fusion reactivity due to the D-T non-Maxwellian ion distribution and its impact on Lawson criterion}, volume={32}, ISSN={1070-664X}, DOI={10.1063/5.0276381}, abstractNote={The non-thermal equilibrium, deviating from the Maxwellian distributions is extremely common in magnetically confined fusion plasmas and has a significant impact on fusion reactivity. Our focus in this contribution is on studying reactant particle distribution functions under steady-state conditions. The kinetic framework resolves steady-state Fokker–Planck solutions to characterize non-thermal ion velocity distributions in three distinct heating configurations: Neutral Beam Injection, three-ion species ion cyclotron resonance heating, and alpha particle heating. Numerical simulations indicate that non-Maxwellian distributions can increase fusion reactivity under appropriate parameter conditions while modifying the Lawson criteria thresholds. This study comprehensively analyzes the effects of kinetic modification on fusion reactivity, identifying quantifiable enhancement pathways for energy gain and providing implementable strategies for advanced reactor design.}, number={9}, journal={Physics of Plasmas}, author={Ye, Yang and Zhang, Wei and Wan, Baonian}, year={2025}, month=sept, pages={092504} }

@article{Garbett_2013, title={Sensitivity of ICF ignition conditions to non-Maxwellian DT fusion reactivity}, volume={59}, rights={© Owned by the authors, published by EDP Sciences, 2013}, ISSN={2100-014X}, DOI={10.1051/epjconf/20135902019}, abstractNote={The hotspot ignition conditions in ICF are determined by considering the power balance between fusion energy deposition and energy loss terms. Uncertainty in any of these terms has potential to modify the ignition conditions, changing the optimum ignition capsule design. This paper considers the impact of changes to the DT fusion reaction rate due to non-thermal ion energy distributions. The DT fusion reactivity has been evaluated for a class of non-Maxwellian distributions representing a perturbation to the tail of a thermal distribution. The resulting reactivity has been used to determine hotspot ignition conditions as a function of the characteristic parameter of the modified distribution.}, journal={EPJ Web of Conferences}, publisher={EDP Sciences}, author={Garbett, W. J.}, year={2013}, pages={02019}, language={en} }

@article{Liu_Li_Yao_Lei_Zhou_Zhu_He_Qiao_2021, title={Enhancement of nuclear reactions via the kinetic Weibel instability in plasmas}, volume={63}, ISSN={0741-3335}, DOI={10.1088/1361-6587/ac2e41}, abstractNote={Nuclear reactions in the plasma environment can be substantially different from those in conventional laboratory non-plasma cases, which have attracted considerable attention in the fields of fusion and astrophysics. To self-consistently model the nuclear reaction process during plasma dynamic evolution, an extended nuclear reaction calculation module is developed and included in two-dimensional particle-in-cell simulations. Through the self-consistent simulations, we systematically show that, apart from the plasma screening, the kinetic Weibel instability (WI) occurring in plasmas also results in significant enhancement of nuclear reactions, where the self-generated magnetic fields play a key role. Specifically, the self-generated magnetic fields in WI deflect ion motions, decreasing the relative velocity, and convert plasma kinetic energy to thermal energy, increasing the ion temperature. The simulation results show that, for the reaction with a sharp resonance peak in the cross section, the reaction product yield is enhanced four times due to the WI. For nuclear reactions that have more prominent resonance peaks in the cross section, like , it is expected that such enhancements can reach up to one or several orders of magnitude.}, number={12}, journal={Plasma Physics and Controlled Fusion}, publisher={IOP Publishing}, author={Liu, Z. Y. and Li, K. and Yao, Y. L. and Lei, Z. and Zhou, C. T. and Zhu, S. P. and He, X. T. and Qiao, B.}, year={2021}, month=nov, pages={125030}, language={en} }

@article{Molvig_Hoffman_Albright_Nelson_Webster_2012, title={Knudsen Layer Reduction of Fusion Reactivity}, volume={109}, DOI={10.1103/PhysRevLett.109.095001}, abstractNote={Knudsen layer losses of tail fuel ions can significantly reduce the fusion reactivity of multi-keV DT in capsules with small fuel ρr; sizable yield reduction can result for small inertial confinement fusion (ICF) capsules. This effect is most pronounced when the distance from a burning DT gas region to a nonreacting or cold wall is comparable to the mean free path of reacting fuel ions. A simplified asymptotic theory of Knudsen layer tail depletion is presented and a nonlocal reduced fusion reactivity model is obtained. Application of the model in simulations of ICF capsule implosion experiments gives calculated yields and ion temperatures that are in much closer agreement with observations than are the results of “nominal” or mixed simulations omitting the model.}, number={9}, journal={Physical Review Letters}, publisher={American Physical Society}, author={Molvig, Kim and Hoffman, Nelson M. and Albright, B. J. and Nelson, Eric M. and Webster, Robert B.}, year={2012}, month=aug, pages={095001} }

@article{Xie_Tan_Luo_Li_Liu_2023, title={Fusion reactivities with drift bi-Maxwellian ion velocity distributions}, volume={65}, ISSN={0741-3335}, DOI={10.1088/1361-6587/acc8f9}, abstractNote={The calculation of fusion reactivity involves a complex six-dimensional integral that takes into account the fusion cross section and velocity distributions of two reactants. However, a more simplified one-dimensional integral form can be useful in certain cases, such as for studying fusion yield or diagnosing ion energy spectra. This simpler form has been derived in a few special cases, such as for a combination of two Maxwellian distributions, a beam-Maxwellian combination, and a beam-target combination, and can greatly reduce computational costs. In this study, it is shown that the reactivity for two drift bi-Maxwellian reactants with different drift velocities, temperatures, and anisotropies can also be reduced to a one-dimensional form, unifying existing derivations into a single expression. This result is used to investigate the potential enhancement of fusion reactivity due to the combination of beam and temperature anisotropies. For relevant parameters in fusion energy, the enhancement factor can be larger than 20%, which is particularly significant for proton-boron (p–B11) fusion, as this factor can have a significant impact on the Lawson fusion gain criteria.}, number={5}, journal={Plasma Physics and Controlled Fusion}, publisher={IOP Publishing}, author={Xie, Huasheng and Tan, Muzhi and Luo, Di and Li, Zhi and Liu, Bing}, year={2023}, month=apr, pages={055019}, language={en} }

@article{Higginson_Link_Schmidt_2019, title={A pairwise nuclear fusion algorithm for weighted particle-in-cell plasma simulations}, volume={388}, ISSN={0021-9991}, DOI={10.1016/j.jcp.2019.03.020}, abstractNote={A pairwise nuclear fusion algorithm for arbitrarily weighted macroparticles in a particle-in-cell simulation is described. The method is benchmarked in situations with like-particles, D(d,n)3He, unlike-particles, D(t,n)4He, thermonuclear plasmas, beam-target fusion, and for large difference in macroparticle weights. Studies of the required number of macroparticles in thermonuclear plasmas show that 100–1000 macroparticles are required to achieve repeatability of yields around 10%, likewise 104–105 for 1%, depending on the fusion interaction and ion temperature.}, journal={Journal of Computational Physics}, author={Higginson, Drew Pitney and Link, Anthony and Schmidt, Andrea}, year={2019}, month=july, pages={439–453} }

@article{Rider_1997, title={Fundamental limitations on plasma fusion systems not in thermodynamic equilibrium}, volume={4}, ISSN={1070-664X}, DOI={10.1063/1.872556}, abstractNote={Analytical Fokker–Planck calculations are used to accurately determine the minimum power that must be recycled in order to maintain a plasma out of thermodynamic equilibrium despite collisions. For virtually all possible types of fusion reactors in which the major particle species are significantly non-Maxwellian or are at radically different mean energies, this minimum recirculating power is substantially larger than the fusion power. Barring the discovery of methods for recycling the power at exceedingly high efficiencies, grossly nonequilibrium reactors will not be able to produce net power.}, number={4}, journal={Physics of Plasmas}, author={Rider, Todd H.}, year={1997}, month=apr, pages={1039–1046} }

@article{McDevitt_Tang_Guo_2017, title={Fast ion transport at a gas-metal interface}, volume={24}, ISSN={1070-664X}, DOI={10.1063/1.4998462}, abstractNote={Fast ion transport and the resulting fusion yield reduction are computed at a gas-metal interface. The extent of fusion yield reduction is observed to depend sensitively on the charge state of the surrounding pusher material and the width of the atomically mixed region. These sensitivities suggest that idealized boundary conditions often implemented at the gas-pusher interface for the purpose of estimating fast ion loss will likely overestimate fusion reactivity reduction in several important limits. In addition, the impact of a spatially complex material interface is investigated by considering a collection of droplets of the pusher material immersed in a DT plasma. It is found that for small Knudsen numbers, the extent of fusion yield reduction scales with the surface area of the material interface. As the Knudsen number is increased, however, the simple surface area scaling is broken, suggesting that hydrodynamic mix has a nontrivial impact on the extent of fast ion losses.}, number={11}, journal={Physics of Plasmas}, author={McDevitt, Christopher J. and Tang, Xian-Zhu and Guo, Zehua}, year={2017}, month=nov, pages={112702} }

@article{Davidovits_Fisch_2014, title={Fusion utility in the Knudsen layer}, volume={21}, ISSN={1070-664X}, DOI={10.1063/1.4895477}, abstractNote={In inertial confinement fusion, the loss of fast ions from the edge of the fusing hot-spot region reduces the reactivity below its Maxwellian value. The loss of fast ions may be pronounced because of the long mean free paths of fast ions, compared with those of thermal ions. We introduce a fusion utility function to demonstrate essential features of this Knudsen layer effect, in both magnetized and unmagnetized cases. The fusion utility concept is also used to evaluate the restoring reactivity in the Knudsen layer by manipulating fast ions in phase space using waves.}, number={9}, journal={Physics of Plasmas}, author={Davidovits, Seth and Fisch, Nathaniel J.}, year={2014}, month=sept, pages={092114} }

@article{Davidovits_Fisch_2016a, title={Sudden Viscous Dissipation of Compressing Turbulence}, volume={116}, ISSN={0031-9007, 1079-7114}, DOI={10.1103/PhysRevLett.116.105004}, number={10}, journal={Physical Review Letters}, author={Davidovits, Seth and Fisch, Nathaniel J.}, year={2016}, month=mar, pages={105004}, language={en} }

@article{Davidovits_Fisch_2016b, title={Compressing turbulence and sudden viscous dissipation with compression-dependent ionization state}, volume={94}, DOI={10.1103/PhysRevE.94.053206}, abstractNote={Turbulent plasma flow, amplified by rapid three-dimensional compression, can be suddenly dissipated under continuing compression. This effect relies on the sensitivity of the plasma viscosity to the temperature, 𝜇∼𝑇5/2. The plasma viscosity is also sensitive to the plasma ionization state. We show that the sudden dissipation phenomenon may be prevented when the plasma ionization state increases during compression, and we demonstrate the regime of net viscosity dependence on compression where sudden dissipation is guaranteed. Additionally, it is shown that, compared to cases with no ionization, ionization during compression is associated with larger increases in turbulent energy and can make the difference between growing and decreasing turbulent energy.}, number={5}, journal={Physical Review E}, publisher={American Physical Society}, author={Davidovits, Seth and Fisch, Nathaniel J.}, year={2016}, month=nov, pages={053206} }

@article{Davidovits_Fisch_2019a, title={Understanding turbulence in compressing plasma as a quasi-EOS}, volume={26}, ISSN={1070-664X}, DOI={10.1063/1.5098790}, abstractNote={Inspired by experimental Z-pinch results, we investigate plasma turbulence undergoing compression. In addition to Z-pinches, plasma turbulence can be compressed in a range of natural and laboratory settings, including inertial fusion experiments and astrophysical molecular clouds. The plasma viscosity, when modeled as described by Braginskii, depends strongly on both temperature and ionization state, giving it the possibility to have a large range of behavior. Here, we highlight the importance of viscous variation in these settings, as well as various insights that can be gained by considering this variation. Included are a “sudden viscous dissipation” effect that leads to a new concept for inertial fusion or X-ray bursts and a bound on turbulent energy behavior under compression. This bound, which was previously applied in inviscid molecular cloud turbulence, is here shown in an application to turbulence that transitions from inviscid to viscous regimes. The task of understanding turbulence under compression can be cast as the process of seeking a “quasi equation of state” for turbulent energy under compression.}, number={6}, journal={Physics of Plasmas}, publisher={American Institute of Physics}, author={Davidovits, Seth and Fisch, Nathaniel J.}, year={2019}, month=june, pages={062709} }

@article{Davidovits_Fisch_2018, title={Bulk hydrodynamic stability and turbulent saturation in compressing hot spots}, volume={25}, ISSN={1070-664X}, DOI={10.1063/1.5026413}, abstractNote={For hot spots compressed at constant velocity, we give a hydrodynamic stability criterion that describes the expected energy behavior of non-radial hydrodynamic motion for different classes of trajectories (in ρR — T space). For a given compression velocity, this criterion depends on ρR, T, and dT/d(ρR) (the trajectory slope) and applies point-wise so that the expected behavior can be determined instantaneously along the trajectory. Among the classes of trajectories are those where the hydromotion is guaranteed to decrease and those where the hydromotion is bounded by a saturated value. We calculate this saturated value and find the compression velocities for which hydromotion may be a substantial fraction of hot-spot energy at burn time. The Lindl (Phys. Plasmas 2, 3933 (1995)] “attractor” trajectory is shown to experience non-radial hydrodynamic energy that grows towards this saturated state. Comparing the saturation value with the available detailed 3D simulation results, we find that the fluctuating velocities in these simulations reach substantial fractions of the saturated value.}, number={4}, journal={Physics of Plasmas}, author={Davidovits, Seth and Fisch, Nathaniel J.}, year={2018}, month=apr, pages={042703} }

@article{Campos_Morgan_2019, title={Self-consistent feedback mechanism for the sudden viscous dissipation of finite-Mach-number compressing turbulence}, volume={99}, DOI={10.1103/PhysRevE.99.013107}, abstractNote={Previous work [Davidovits and Fisch, Phys. Rev. Lett. 116, 105004 (2016)] demonstrated that the compression of a turbulent field can lead to a sudden viscous dissipation of turbulent kinetic energy (TKE), and that paper suggested this mechanism could potentially be used to design new fast-ignition schemes for inertial confinement fusion (ICF). We expand on previous work by accounting for finite Mach numbers, rather than relying on a zero-Mach-limit assumption as previously done. The finite-Mach-number formulation is necessary to capture a self-consistent feedback mechanism in which dissipated TKE increases the temperature of the system, which in turn modifies the viscosity and thus the TKE dissipation itself. Direct numerical simulations with a tenth-order accurate Padé scheme were carried out to analyze this self-consistent feedback loop for compressing turbulence. Results show that, for finite Mach numbers, the sudden viscous dissipation of TKE still occurs, for both the solenoidal and dilatational turbulent fields. As the domain is compressed, oscillations in dilatational TKE are encountered due to the highly oscillatory nature of the pressure dilatation. An analysis of the source terms for the internal energy shows that the mechanical-work term dominates the viscous turbulent dissipation. As a result, the effect of the suddenly dissipated TKE on temperature is minimal for the Mach numbers tested. Moreover, an analytical expression is derived that confirms the dissipated TKE does not significantly alter the temperature evolution for low Mach numbers, regardless of compression speed. The self-consistent feedback mechanism is thus quite weak for subsonic turbulence, which could limit its applicability for ICF.}, number={1}, journal={Physical Review E}, publisher={American Physical Society}, author={Campos, Alejandro and Morgan, Brandon E.}, year={2019}, month=jan, pages={013107} }

@article{Davidovits_Fisch_2019b, title={Viscous dissipation in two-dimensional compression of turbulence}, volume={26}, ISSN={1070-664X}, DOI={10.1063/1.5111961}, abstractNote={Nonradial hydrodynamic flow can be generated or amplified during plasma compression by various mechanisms, including the compression itself. In certain circumstances, the plasma may reach a viscous state; for example, in compression experiments seeking fusion, the fuel plasma may reach a viscous state late in the compression due in part to the rising fuel temperature. Here, we consider viscous dissipation of nonradial flow in the case of initially isotropic, three-dimensional (3D), turbulent flow fields compressed at constant velocity in two dimensions. Prior work in the case of 3D compressions has shown the possibility of effective viscous dissipation of nonradial flow under compression. We show that, theoretically, complete viscous dissipation of the nonradial flow should still occur in the 2D case when the plasma heating is adiabatic and the viscosity has the (strong) Braginskii temperature dependence (μ∼T5/2). However, in the general case, the amount of compression required is very large even for modest initial Reynolds numbers, with the compression reaching an intermediate state dominated by variations only in the noncompressed direction. We show that both the nonlinearity and boundary conditions can play important roles in setting the characteristics and ease of the viscous dissipation.}, number={8}, journal={Physics of Plasmas}, author={Davidovits, Seth and Fisch, Nathaniel J.}, year={2019}, month=aug, pages={082702} }

@article{Li_Davidovits_2024, title={Microphysics of shock-grain interaction for inertial confinement fusion ablators in a fluid approach}, volume={110}, DOI={10.1103/PhysRevE.110.035206}, abstractNote={Ablator materials used for inertial confinement fusion, such as high-density carbon (HDC) and beryllium, have grain structure which may lead to small-scale density nonuniformity and the generation of perturbations when the materials are shocked and compressed. Here, we use a combination of a linear theory of shock interaction with density nonuniformity [Velikovich et al., Phys. Plasmas 14, 072706 (2007)] and numerical simulations to study shock interaction with a model representation of HDC grains. While the shock-grain interaction is nonlinear, the linear theory shows some key features of the shock-grain interaction, which also hold for the (nonlinear) simulations. The postshock perturbations are made up of sonic reflections off of grain boundaries and vorticity deposition along them, with the latter dominating the perturbed energy content. The mean (per mass) postshock perturbed kinetic energy decreases with increasing grain size, but energy will be deposited at increasing spatial scale. From the perspective of the postshock perturbed energy, the detailed linear theory largely supports a proposed method [S. Davidovits et al., Phys. Plasmas 29, 112708 (2022)] for deresolving the grains (in a similar grains model) that treats the grains statistically. Our simulation results highlight the influence of thermal conduction on the perturbation dynamics at grain scales.}, number={3}, journal={Physical Review E}, publisher={American Physical Society}, author={Li, G. J. and Davidovits, S.}, year={2024}, month=sept, pages={035206} }

@article{Albright_Molvig_Huang_Simakov_Dodd_Hoffman_Kagan_Schmit_2013, title={Revised Knudsen-layer reduction of fusion reactivity}, volume={20}, ISSN={1070-664X}, DOI={10.1063/1.4833639}, abstractNote={Recent work by Molvig et al. [Phys. Rev. Lett. 109, 095001 (2012)] examined how fusion reactivity may be reduced from losses of fast ions in finite assemblies of fuel. In this paper, this problem is revisited with the addition of an asymptotic boundary-layer treatment of ion kinetic losses. This boundary solution, reminiscent of the classical Milne problem from linear transport theory, obtains a free-streaming limit of fast ion losses near the boundary, where the diffusion approximation is invalid. Thermonuclear reaction rates have been obtained for the ion distribution functions predicted by this improved model. It is found that while Molvig’s “Knudsen distribution function” bounds from above the magnitude of the reactivity reduction, this more accurate treatment leads to less dramatic losses of tail ions and associated reduction of thermonuclear reaction rates for finite fuel volumes.}, number={12}, journal={Physics of Plasmas}, author={Albright, B. J. and Molvig, Kim and Huang, C.-K. and Simakov, A. N. and Dodd, E. S. and Hoffman, N. M. and Kagan, G. and Schmit, P. F.}, year={2013}, month=dec, pages={122705} }

@article{Bosch_Hale_1992, title={Improved formulas for fusion cross-sections and thermal reactivities}, volume={32}, ISSN={0029-5515}, DOI={10.1088/0029-5515/32/4/I07}, abstractNote={For interpreting fusion rate measurements in present fusion experiments and predicting the fusion performance of future devices or of d-t experiments in present devices, it is important to know the fusion cross-sections as precisely as possible. Usually, it is not measured data that are used, but parametrizations of the cross-section as a function of the ion energy and parametrizations of the Maxwellian reactivity as a function of the ion-temperature. Since the publication of the parametrizations now in use, new measurements have been made and evaluations of the measured data have been improved by applying R-matrix theory. The authors show that the old parametrizations no longer adequately represent the experimental data and present new parametrizations based on R-matrix calculations for fusion cross-sections and Maxwellian reactivities for the reactions D(d,n)3He, D(d,p)T, T(d,n)4He and 3He(d,p)4He}, number={4}, journal={Nuclear Fusion}, author={Bosch, H.-S. and Hale, G. M.}, year={1992}, month=apr, pages={611}, language={en} }

@article{Adrian_et_2025, title={Constraining the 3He + 3He Gamow energy probed in high energy density plasmas at the National Ignition Facility}, volume={32}, ISSN={1070-664X}, DOI={10.1063/5.0233437}, abstractNote={Polar-direct-drive implosions at the National Ignition Facility generated large
plasma volumes to study the 3He + 3He fusion reaction. The
ion temperature, which determines the Gamow peak energy, was constrained by
isolating the thermal contribution to the D3He-proton spectral width
in a 3He plasma doped with deuterium. X-ray penumbral imaging was
used to measure electron temperature, density, and hotspot volume, which was
subsequently used to model the spectral broadening from plasma stopping power.
Results showed 30% of the D3He-proton spectral width was due to
stopping power, with residual flows contributing  ≈10%. The 3He
temperature was determined as  T3He = 12.4 ± 3.2 keV,
corresponding to a Gamow energy of 95 ± 14 keV. These experiments achieved the
lowest Gamow energy to date for studying 3He + 3He fusion
in high energy density plasma, approaching conditions in the Sun.}, number={2}, journal={Physics of Plasmas}, author={Adrian, P. J. and Bachmann, B. and Casey, D. T. and Craxton, R. S. and Garbett, W. and Johnson, M. Gatu and Hartouni, E. and Hohenberger, M. and Holunga, D. and Kabadi, N. V. and others}, year={2025}, month=feb, pages={022704} }

@article{Casey_et_2017, title={Thermonuclear reactions probed at stellar-core conditions with laser-based inertial-confinement fusion}, volume={13}, rights={2017 Springer Nature Limited}, ISSN={1745-2481}, DOI={10.1038/nphys4220}, abstractNote={Stars are giant thermonuclear plasma furnaces that slowly fuse the lighter elements in the universe into heavier elements, releasing energy, and generating the pressure required to prevent collapse. To understand stars, we must rely on nuclear reaction rate data obtained, up to now, under conditions very different from those of stellar cores. Here we show thermonuclear measurements of the 2H(d, n)3He and 3H(t,2n)4He S-factors at a range of densities (1.2–16?g?cm−3) and temperatures (2.1–5.4?keV) that allow us to test the conditions of the hydrogen-burning phase of main-sequence stars. The relevant conditions are created using inertial-confinement fusion implosions at the National Ignition Facility. Our data agree within uncertainty with previous accelerator-based measurements and establish this approach for future experiments to measure other reactions and to test plasma-nuclear effects present in stellar interiors, such as plasma electron screening, directly in the environments where they occur.}, number={12}, journal={Nature Physics}, publisher={Nature Publishing Group}, author={Casey, D. T. and Sayre, D. B. and Brune, C. R. and Smalyuk, V. A. and Weber, C. R. and Tipton, R. E. and Pino, J. E. and Grim, G. P. and Remington, B. A. and Dearborn, D. and others}, year={2017}, month=dec, pages={1227–1231}, language={en} }

@article{Casey_et._2023, title={Towards the first plasma-electron screening experiment}, volume={10}, ISSN={2296-424X}, url={https://www.frontiersin.org/journals/physics/articles/10.3389/fphy.2022.1057603/full}, DOI={10.3389/fphy.2022.1057603}, abstractNote={The enhancement of fusion reaction rates in a thermonuclear plasma by electron screening of the Coulomb barrier is an important plasma-nuclear effect that is present in stellar models but has not been experimentally observed. Experiments using inertial confinement fusion (ICF) implosions may provide a unique opportunity to observe this important plasma-nuclear effect. Herein, we show that experiments at the National Ignition Facility (NIF) have reached the relevant physical regime with respect to the density and temperature conditions but the estimated impacts of plasma screening on nuclear reaction rates are currently too small and need to be increased to lower the expected measurement uncertainty. Detailed radiation hydrodynamics simulations show that practical target changes, like adding readily available high-Z gases, and significantly slowing the inflight implosion velocity while maintaining inflight kinetic energy, might be able to push these conditions to those where plasma screening effects may be measurable. We also perform synthetic data exercises to help understand where the anticipated experimental uncertainties will become important. But challenges remain such as detectability of the reaction products, non-thermal plasma effects, species separation, and impacts of spatial and temporal gradients. This work lays the foundation for future efforts to develop an important platform capable of the first plasma electron screening observation.}, journal={Frontiers in Physics}, publisher={Frontiers}, author={Casey, Daniel T. and Weber, Chris R. and Zylstra, Alex B. and Cerjan, Charlie J. and Hartouni, Ed and Hohenberger, Matthias and Divol, Laurent and Dearborn, David S. and Kabadi, Neel and Lahmann, Brandon and Gatu Johnson, Maria and Frenje, Johan A.}, year={2023}, month=jan, language={English} }

@article{ICF_collab_2024, title={Achievement of Target Gain Larger than Unity in an Inertial Fusion Experiment}, volume={132}, DOI={10.1103/PhysRevLett.132.065102}, abstractNote={On December 5, 2022, an indirect drive fusion implosion on the National Ignition Facility (NIF) achieved a target gain Gtarget of 1.5. This is the first laboratory demonstration of exceeding “scientific breakeven” (or Gtarget>1) where 2.05 MJ of 351 nm laser light produced 3.1 MJ of total fusion yield, a result which significantly exceeds the Lawson criterion for fusion ignition as reported in a previous NIF implosion [H. Abu-Shawareb et al. (Indirect Drive ICF Collaboration), Phys. Rev. Lett. 129, 075001 (2022)]. This achievement is the culmination of more than five decades of research and gives proof that laboratory fusion, based on fundamental physics principles, is possible. This Letter reports on the target, laser, design, and experimental advancements that led to this result.}, number={6}, journal={Physical Review Letters}, publisher={American Physical Society}, author={Abu-Shawareb, H. and Acree, R. and Adams, P. and Adams, J. and Addis, B. and Aden, R. and Adrian, P. and Afeyan, B. B. and Aggleton, M. and Aghaian, L.  and others},
 collaboration = {The Indirect Drive ICF Collaboration}, year={2008}, month=oct, pages={102707} }

@article{Kritcher_et_2024, title={Design of the first fusion experiment to achieve target energy gain $G>1$}, volume={109}, DOI={10.1103/PhysRevE.109.025204}, abstractNote={In this work we present the design of the first controlled fusion laboratory experiment to reach target gain G>1 N221204 (5 December 2022) [Phys. Rev. Lett. 132, 065102 (2024)], performed at the National Ignition Facility, where the fusion energy produced (3.15 MJ) exceeded the amount of laser energy required to drive the target (2.05 MJ). Following the demonstration of ignition according to the Lawson criterion N210808, experiments were impacted by nonideal experimental fielding conditions, such as increased (known) target defects that seeded hydrodynamic instabilities or unintentional low-mode asymmetries from nonuniformities in the target or laser delivery, which led to reduced fusion yields less than 1 MJ. This Letter details design changes, including using an extended higher-energy laser pulse to drive a thicker high-density carbon (also known as diamond) capsule, that led to increased fusion energy output compared to N210808 as well as improved robustness for achieving high fusion energies (greater than 1 MJ) in the presence of significant low-mode asymmetries. For this design, the burnup fraction of the deuterium and tritium (DT) fuel was increased (approximately 4% fuel burnup and a target gain of approximately 1.5 compared to approximately 2% fuel burnup and target gain approximately 0.7 for N210808) as a result of increased total (DT plus capsule) areal density at maximum compression compared to N210808. Radiation-hydrodynamic simulations of this design predicted achieving target gain greater than 1 and also the magnitude of increase in fusion energy produced compared to N210808. The plasma conditions and hotspot power balance (fusion power produced vs input power and power losses) using these simulations are presented. Since the drafting of this manuscript, the results of this paper have been replicated and exceeded (N230729) in this design, together with a higher-quality diamond capsule, setting a new record of approximately 3.88MJ of fusion energy and fusion energy target gain of approximately 1.9.}, number={2}, journal={Physical Review E}, publisher={American Physical Society}, author={Kritcher, A. L. and Zylstra, A. B. and Weber, C. R. and Hurricane, O. A. and Callahan, D. A. and Clark, D. S. and Divol, L. and Hinkel, D. E. and Humbird, K. and Jones, O. and others}, year={2024}, month=feb, pages={025204} }

@article{Albright_et_2022, title={Experimental quantification of the impact of heterogeneous mix on thermonuclear burn}, volume={29}, ISSN={1070-664X}, DOI={10.1063/5.0082344}, abstractNote={In inertial confinement fusion, deuterium–tritium (DT) fuel is brought to densities and temperatures where fusion ignition occurs. However, mixing of the ablator material into the fuel may prevent ignition by diluting and cooling the fuel. MARBLE experiments at the National Ignition Facility provide new insight into how mixing affects thermonuclear burn. These experiments use laser-driven capsules containing deuterated plastic foam and tritium gas. Embedded within the foam are voids of known sizes and locations, which control the degree of heterogeneity of the fuel. Initially, the reactants are separated, with tritium concentrated in the voids and deuterium in the foam. During the implosion, mixing occurs between the foam and gas materials, leading to DT fusion reactions in the mixed region. Here, it is shown that by measuring the ratios of DT and deuterium–deuterium neutron yields for different macropore sizes and gas compositions, the effects of mix heterogeneity on thermonuclear burn may be quantified, supporting an improved understanding of these effects.}, number={2}, journal={Physics of Plasmas}, author={Albright, B. J. and Murphy, T. J. and Haines, B. M. and Douglas, M. R. and Cooley, J. H. and Day, T. H. and Denissen, N. A. and Di Stefano, C. and Donovan, P. and Edwards, S. L. and others}, year={2022}, month=feb, pages={022702} }

@article{Mannion_et._2023, title={Evidence of non-Maxwellian ion velocity distributions in spherical shock-driven implosions}, volume={108}, DOI={10.1103/PhysRevE.108.035201}, abstractNote={The ion velocity distribution functions of thermonuclear plasmas generated by spherical laser direct drive implosions are studied using deuterium-tritium (DT) and deuterium-deuterium (DD) fusion neutron energy spectrum measurements. A hydrodynamic Maxwellian plasma model accurately describes measurements made from lower temperature (<10 keV), hydrodynamiclike plasmas, but is insufficient to describe measurements made from higher temperature more kineticlike plasmas. The high temperature measurements are more consistent with Vlasov-Fokker-Planck (VFP) simulation results which predict the presence of a bimodal plasma ion velocity distribution near peak neutron production. These measurements provide direct experimental evidence of non-Maxwellian ion velocity distributions in spherical shock driven implosions and provide useful data for benchmarking kinetic VFP simulations.}, number={3}, journal={Physical Review E}, publisher={American Physical Society}, author={Mannion, O. M. and Taitano, W. T. and Appelbe, B. D. and Crilly, A. J. and Forrest, C. J. and Glebov, V. Yu. and Knauer, J. P. and McKenty, P. W. and Mohamed, Z. L. and Stoeckl, C. and others}, year={2023}, month=sep, pages={035201} }

@article{Petschek_Henderson_1979, title={Influence of high-energy ion loss on DT reaction rate in laser fusion pellets}, volume={19}, ISSN={0029-5515}, DOI={10.1088/0029-5515/19/12/012}, abstractNote={Because of the longer mean free path of highenergy ions, they will be preferentially lost from small pellets containing thermonuclear reactants. This effect has been calculated and, in the most extreme case calculated, a factor-of about-four reduction of the reaction rate in DT from the Maxwell average rate at the same mean ion kinetic energy is found.}, number={12}, journal={Nuclear Fusion}, author={Petschek, A. G. and Henderson, D. B.}, year={1979}, month=dec, pages={1678}, language={en} }

@article{Jeet_et_2023, title={Observations of multi-ion physics and kinetic effects in a surrogate to the solar CNO reactions}, volume={49}, ISSN={1574-1818}, DOI={10.1016/j.hedp.2023.101066}, abstractNote={The ‘CNO process’ occurs in heavier stars with finite metallicity in which hydrogen burning is catalyzed in the presence of 12C. These reactions are more strongly dependent on temperature than the pp cycle reactions, and thus the CNO cycle dominates only in massive stars. For these types of reactions to be studied at ICF facilities such as OMEGA, an implosion platform using heavier nuclei in the fuel and capable of creating ion temperatures on the order of at least 20 keV is required. A potential route to reach these conditions is to take advantage of kinetic effects in low-convergence shock-driven ‘exploding pusher’ implosions. In this experiment, shots were conducted at the OMEGA laser facility using the surrogate reaction 13C + D. Its cross section is substantially higher than the actual astrophysical CNO reactions. The yield of this reaction in these implosions was much lower than expected. Physical explanations are discussed, with significant species stratification the likely explanation.}, journal={High Energy Density Physics}, author={Jeet, J. and Zylstra, A. B. and Gatu Johnson, M. and Kabadi, N. V. and Adrian, P. and Forrest, C. and Glebov, V.}, year={2023}, month=dec, pages={101066} }

@article{Davidovits_et_2022, title={Turbulence generation by shock interaction with a highly nonuniform medium}, volume={105}, DOI={10.1103/PhysRevE.105.065206}, abstractNote={An initially planar shock wave propagating into a medium of nonuniform density will be perturbed, leading to the generation of postshock velocity perturbations. Using numerical simulations we study this phenomenon in the case of highly nonuniform density (order-unity normalized variance, 𝜎𝜌/‾‾‾𝜌∼1) and strong shocks (shock Mach numbers ‾‾‾𝑀𝑠≳10). This leads to a highly disrupted shock and a turbulent postshock flow. We simulate this interaction for a range of shock drives and initial density configurations meant to mimic those which might be presently achieved in experiments. Theoretical considerations lead to scaling relations, which are found to reasonably predict the postshock turbulence properties. The turbulent velocity dispersion and turbulent Mach number are found to depend on the preshock density dispersion and shock speed in a manner consistent with the linear Richtymer-Meshkov instability prediction. We also show a dependence of the turbulence generation on the scale of density perturbations. The postshock pressure and density, which can be substantially reduced relative to the unperturbed case, are found to be reasonably predicted by a simplified analysis that treats the extended shock transition region as a single normal shock.}, number={6}, journal={Physical Review E}, publisher={American Physical Society}, author={Davidovits, Seth and Federrath, Christoph and Teyssier, Romain and Raman, Kumar S. and Collins, David C. and Nagel, Sabrina R.}, year={2022}, month=jun, pages={065206} }

@article{Haines_et_2023, title={The dynamics, mixing, and thermonuclear burn of compressed foams with varied gas fills}, volume={30}, ISSN={1070-664X}, DOI={10.1063/5.0154600}, abstractNote={Inertial confinement fusion (ICF) implosions involve highly coupled physics and complex hydrodynamics that are challenging to model computationally. Due to the sensitivity of such implosions to small features, detailed simulations require accurate accounting of the geometry and dimensionality of the initial conditions, including capsule defects and engineering features such as fill tubes used to insert gas into the capsule, yet this is computationally prohibitive. It is therefore difficult to evaluate whether discrepancies between the simulation and experiment arise from inadequate fidelity to the capsule geometry and drive conditions, uncertainties in physical data used by simulations, or inadequate physics. We present results from detailed high-resolution three-dimensional simulations of ICF implosions performed as part of the MARBLE campaign on the National Ignition Facility [Albright et al., Phys. Plasmas 29, 022702 (2022)]. These experiments are foam-filled separated-reactant experiments, where deuterons reside in the foam and tritons reside in the capsule gas fill and deuterium–tritium (DT) fusion reactions only occur in the presence of mixing between these materials. Material mixing in these experiments is primarily seeded by shock interaction with the complex geometry of the foam and gas fill, which induces the Richtmyer–Meshkov instability. We compare results for experiments with two different gas fills (ArT and HT), which lead to significant differences in the hydrodynamic and thermodynamic developments of the materials in the implosion. Our simulation results show generally good agreement with experiments and demonstrate a substantial impact of hydrodynamic flows on measured ion temperatures. The results suggest that viscosity, which was not included in our simulations, is the most important unmodeled physics and qualitatively explains the few discrepancies between the simulation and experiment. The results also suggest that the hydrodynamic treatment of shocks is inadequate to predict the heating and yield produced during shock flash, when the shock converges at the center of the implosion. Alternatively, underestimation of the level of radiative preheat from the shock front could explain many of the differences between the experiment and simulation. Nevertheless, simulations are able to reproduce many experimental observables within the level of experimental reproducibility, including most yields, time-resolved X-ray self-emission images, and an increase in burn-weighted ion temperature and neutron down-scattered ratio in the line of sight that includes a jet seeded by the glue spot that joins capsule hemispheres.}, number={7}, journal={Physics of Plasmas}, author={Haines, Brian M. and Murphy, T. J. and Olson, R. E. and Kim, Y. and Albright, B. J. and Appelbe, B. and Day, T. H. and Gunderson, M. A. and Hamilton, C. E. and Morrow, T. and Patterson, B. M.}, year={2023}, month=jul, pages={072705} }

@article{Murphy_et_2021, title={Results from single-shock Marble experiments studying thermonuclear burn in the presence of heterogeneous mix on the National Ignition Facility}, volume={38}, ISSN={1574-1818}, DOI={10.1016/j.hedp.2021.100929}, abstractNote={The Marble campaign on the National Ignition Facility investigates the effect of heterogeneous mix on thermonuclear burn for comparison to a probability distribution function (PDF) burn model. Marble utilizes plastic capsules filled with deuterated plastic foam and a fill gas containing tritium. As the capsules implode, the deuterium in the foam mixes with the tritium gas, and DT neutrons are produced as the shocks compress and heat the mixture. The yield of DT neutrons is dependent on the uniformity of the mix, with more heterogeneous mix producing fewer neutrons. In Marble, the heterogeneity of the mix is controlled by varying the diameter of voids introduced into the foam. The first NIF Marble campaign has been executed in which the Marble capsules were indirectly driven with a single strong shock using NIF hohlraums. The experiments produce a low-convergence, high-ion-temperature implosion. The ratio of DT to DD neutron yield is largely consistent with uniform atomic mix for fine-pore foam, and increases slightly with void diameter, contrary to 1D simulations using the PDF burn model. Recent 3D high-resolution simulations of similar experiments performed on the Omega Laser Facility suggest an explanation.}, journal={High Energy Density Physics}, author={Murphy, Thomas J. and Albright, B. J. and Douglas, M. R. and Cardenas, T. and Cooley, J. H. and Day, T. H. and Denissen, N. A. and Gore, R. A. and Gunderson, M. A. and Haack, J. R. and others}, year={2021}, month=mar, pages={100929} }

@article{Fetsch_Foster_Fisch_2023, title={Temperature separation under compression of moderately coupled plasma}, volume={89}, ISSN={0022-3778, 1469-7807}, DOI={10.1017/S0022377823000776}, abstractNote={In moderately coupled plasmas, a significant fraction of the internal energy resides in electric fields. As these plasmas are heated or compressed, the shifting partition of energy between particles and fields leads to surprising effects, particularly when ions and electrons have different temperatures. In this work, quasi-equations of state (quasi-EOS) are derived for two-temperature moderately coupled plasma in a thermodynamic framework and expressed in a simple form. These quasi-EOS readily yield expressions for correlation heating, in which heating of the electrons causes a rapid increase in ion temperature even in the absence of collisional energy exchange between species. It is also shown that, remarkably, compression of moderately coupled plasma drives a temperature difference between electrons and ions, even when the species start at equal temperatures. These additional channels for ion heating may be relevant in designing ignition schemes for inertial confinement fusion.}, number={5}, journal={Journal of Plasma Physics}, author={Fetsch, H. and Foster, T. E. and Fisch, N. J.}, year={2023}, month=oct, pages={905890510}, language={en} }

@article{Foster_Fetsch_Fisch_2023, title={Fast correlation heating in moderately coupled electron–ion plasmas}, volume={89}, ISSN={0022-3778, 1469-7807}, DOI={10.1017/S0022377823000922}, abstractNote={If the electrons in a plasma are suddenly heated, the resulting change in Debye shielding causes the ion kinetic energy to quickly increase. For the first time, this correlation heating, which is much faster than collisional energy exchange, is rigorously derived for a moderately coupled, electron–ion plasma. The electron–ion mass ratio is taken to be the smallest parameter in the Bogoliubov–Born–Green–Kirkwood–Yvon hierarchy, smaller even than the reciprocal of the plasma parameter. This ordering differs from conventional kinetic theory by making the electron collision rates faster than the ion plasma frequency, which allows stronger coupling and makes the ion heating a function only of the total energy supplied to the electrons. The calculation uses known formulae for correlations in a two-temperature plasma, for which a new, elementary derivation is presented. Suprathermal ions may be created more rapidly by this mechanism than by ion–electron Coulomb collisions. This means that the use of a femtosecond laser pulse could potentially help to achieve ignition in certain fast ignition approaches to inertial confinement fusion.}, number={5}, journal={Journal of Plasma Physics}, author={Foster, Thomas E. and Fetsch, Henry and Fisch, Nathaniel J.}, year={2023}, month=oct, pages={905890506}, language={en} }

@article{Clarke_1980, title={Hot-ion-mode ignition in a tokamak reactor}, volume={20}, ISSN={0029-5515}, DOI={10.1088/0029-5515/20/5/005}, abstractNote={Presently observed scaling laws in tokamak experiments allow ignition of tokamak reactors with nre possibly as low as 4.5 × 1013 cm−3·s. These reactors operate in the hot-ion ignition mode with Ti>Te and are a direct extension of the hot-ion-mode operation observed in present tokamaks and expected in TFTR. They require MHD stability similar to conventional tokamak reactors and microstability at collision-ality ten times lower than that observed on PLT. All physics issues associated with hot-ion ignited reactors, short of complete ignition, can be addressed in existing facilities and TFTR.}, number={5}, journal={Nuclear Fusion}, author={Clarke, J.F.}, year={1980}, month=may, pages={563}, language={en} }

@article{Fisch_Herrmann_1994, title={Utility of extracting alpha particle energy by waves}, volume={34}, ISSN={0029-5515}, DOI={10.1088/0029-5515/34/12/I01}, abstractNote={The utility of extracting alpha particle power, and then diverting this power to fast fuel ions, is investigated. As power is diverted to fast ions and then to ions, a number of effects come into play, as the relative amounts of pressure taken up by electrons, fuel ions and fast alpha particles shift. In addition, if the alpha particle power is diverted to fast fuel ions, there is an enhanced fusion reactivity because of the non-thermal component of the ion distribution. Some useful expressions for describing these effects are derived, and it is shown that fusion reactors with power density about twice what otherwise might be obtained can be contemplated, so long as a substantial amount of the alpha particle power can be diverted. Interestingly, in this mode of operation, once the electron heat is sufficiently confined, further improvement in confinement is actually not desirable. A similar improvement in fusion power density can be obtained for advanced fuel mixtures such as D-3He, where the power of both the energetic alpha particles and the energetic protons might be diverted advantageously}, number={12}, journal={Nuclear Fusion}, author={Fisch, N. J. and Herrmann, M. C.}, year={1994}, month=dec, pages={1541}, language={en} }

@article{Fisch_Herrmann_1999, title={A tutorial on -channelling}, volume={41}, ISSN={0741-3335}, DOI={10.1088/0741-3335/41/3A/015}, abstractNote={One of the more ambitious uses of intense microwaves in tokamaks or in other magnetic confinement deuterium-tritium (DT) fusion devices would be to divert power from energetic -particles to waves. This so-called `-channelling’ would be a large step towards achieving economical fusion power. The intense waves, amplified by the substantial free energy in the -particles, damp on fuel ions, resulting in a hot ion mode, doubling the fusion power of the reactor at the same confined pressure. If the waves damp preferentially on electrons or ions travelling in one direction, current can be driven. This tutorial explains the key concepts and recent advances that lead us to believe in the plausibility of such an effect, at the same time showing how experiments to date give us a measure of confidence in both the simulations themselves, the underlying physical assumptions and ultimately the reasonableness of the application of these ideas to -channelling in a tokamak reactor.}, number={3A}, journal={Plasma Physics and Controlled Fusion}, author={Fisch, Nathaniel J. and Herrmann, Mark C.}, year={1999}, month=mar, pages={A221}, language={en} }

@article{Fisch_Rax_1992, title={Interaction of energetic alpha particles with intense lower hybrid waves}, volume={69}, DOI={10.1103/PhysRevLett.69.612}, abstractNote={Lower hybrid waves are a demonstrated, continuous means of driving toroidal current in a tokamak. When these waves propagate in a tokamak fusion reactor, in which there are energetic α particles, there are conditions under which the α particles do not appreciably damp, and may even amplify, the wave, thereby enhancing the current-drive effect. Waves traveling in one poloidal direction, in addition to being directed in one toroidal direction, are shown to be the most efficient drivers of current in the presence of the energetic α particles.}, number={4}, journal={Physical Review Letters}, publisher={American Physical Society}, author={Fisch, Nathaniel J. and Rax, Jean-Marcel}, year={1992}, month=july, pages={612–615} }

@article{Furth_Jassby_1974, title={Power Amplification Conditions for Fusion-Reactor Plasmas Heated by Reacting Ion Beams}, volume={32}, DOI={10.1103/PhysRevLett.32.1176}, abstractNote={The plasma energy confinement time required to achieve a practically useful power amplification factor 𝑄𝐸 can be reduced substantially if the plasma is heated by an energetic ion beam that reacts with the bulk ions as it thermalizes. The contribution to 𝑄𝐸 arising from the beam-plasma fusion reactions can be maximized by maintaining the post-injection energy of the beam ions near an optimum value, for example, by means of slow magnetic compression.}, number={21}, journal={Physical Review Letters}, publisher={American Physical Society}, author={Furth, H. P. and Jassby, D. L.}, year={1974}, month=may, pages={1176–1179} }

@article{Kroupp_et_2018, title={Turbulent stagnation in a $Z$-pinch plasma}, volume={97}, DOI={10.1103/PhysRevE.97.013202}, abstractNote={The ion kinetic energy in a stagnating plasma was previously determined by Kroupp et al. [Phys. Rev. Lett. 107, 105001 (2011)] from Doppler-dominated line shapes augmented by measurements of plasma properties and assuming a uniform-plasma model. Notably, the energy was found to be dominantly stored in hydrodynamic flow. Here we advance a new description of this stagnation as supersonically turbulent. Such turbulence implies a nonuniform density distribution. We demonstrate how to reanalyze the spectroscopic data consistent with the turbulent picture and show that this leads to better concordance of the overconstrained spectroscopic measurements, while also substantially lowering the inferred mean density.}, number={1}, journal={Physical Review E}, publisher={American Physical Society}, author={Kroupp, E. and Stambulchik, E. and Starobinets, A. and Osin, D. and Fisher, V. I. and Alumot, D. and Maron, Y. and Davidovits, S. and Fisch, N. J. and Fruchtman, A.}, year={2018}, month=jan, pages={013202} }

@article{Reichelt_Petrasso_Li_2024, title={Effects of alpha-ion stopping on ignition and ignition criteria in inertial confinement fusion experiments}, volume={31}, ISSN={1070-664X}, DOI={10.1063/5.0180544}, abstractNote={With the advent of ignited plasmas at the National Ignition Facility (NIF), alpha physics has become a driving factor in theoretical understanding and experimental behavior. In this communication, we explore aspects of direct alpha-ion heating through comparison of the consequences from the one-fluid and two-fluid models in the hydrodynamic approach. We show that the case with all alpha energy deposited in electrons raises the ignition criteria by    ∼ 4   keV or    ∼ 0.2     g / cm 2 in the hotspot relative to the case with all alpha energy deposited in ions. In the case of the recently ignited NIF implosion, 30% of the 3.5 MeV α energy is deposited into the DT fuel ions, for which there is negligible difference between the one-fluid and two-fluid ignition criteria. However, changes in the ion stopping fraction through profile effects and alternate stopping power models could lead to ignition curve shifts of    ∼ 1   keV.}, number={1}, journal={Physics of Plasmas}, author={Reichelt, Benjamin L. and Petrasso, Richard D. and Li, Chikang}, year={2024}, month=jan, pages={010702} }

@article{Du_Kang_Zou_Liu_Deng_Ge_Dai_Cai_Zhu_2024, title={Modeling of the non-Maxwellian response of DT plasmas to alpha particle transport in inertial confinement fusion (ICF) hotspot}, volume={31}, ISSN={1070-664X}, DOI={10.1063/5.0179526}, abstractNote={In the alpha particle transport in ICF hotspot, previous models focus mainly on how the incident particles lose their energy but lost sight of how the target particles will respond to this lost energy. In this paper, we developed a novel single-scattering model based on the Monte Carlo method, which abandons the stopping-power and models every single-scattering event in the alpha particle life. It enables to describe both the energy stopping of the incident alpha particle and the target particles response to the collisions. With this model, it shows that the target DT-ions at the ICF hotspot boundary will be non-Maxwellian distributed after colliding with the high-energy alpha particles, which refers to a much higher fusion reactivity compared with a Maxwellian one. At the same time, this model gives a longer and dispersed alpha particle range in hotspot plasmas and suggests that the traditionally used stopping power models would overestimate the stopping ability of the target particles.}, number={1}, journal={Physics of Plasmas}, author={Du, Bao and Kang, Dongguo and Zou, Shiyang and Liu, Chang and Deng, Luan and Ge, Fengjun and Dai, Zhensheng and Cai, Hongbo and Zhu, Shaoping}, year={2024}, month=jan, pages={012706} }

@article{Li_Petrasso_1993a, title={Charged-particle stopping powers in inertial confinement fusion plasmas}, volume={70}, DOI={10.1103/PhysRevLett.70.3059}, abstractNote={The effects of large-angle scattering, important for plasmas for which the Coulomb logarithm is of order 1, have been properly treated in calculating the range (R) and the ρR (the fuel-areal density) of inertial confinement fusion plasmas. This new calculation, which also includes the important effects of plasma ion stopping, collective plasma oscillations, and quantum effects, leads to an accurate estimate, not just an upper limit of ρR. For example, 3.5 MeV α’s from D-T fusion reactions are found to directly deposit ≃47% of their energy into 20 keV deuterons and tritons. Consequently the α range (R) and ρR are reduced by about 60% from the case of pure electron stopping.}, number={20}, journal={Physical Review Letters}, publisher={American Physical Society}, author={Li, Chi-Kang and Petrasso, Richard D.}, year={1993}, month=may, pages={3059–3062} }

@article{Li_Petrasso_1993b, title={Fokker-Planck equation for moderately coupled plasmas}, volume={70}, rights={http://link.aps.org/licenses/aps-default-license}, ISSN={0031-9007}, DOI={10.1103/PhysRevLett.70.3063}, number={20}, journal={Physical Review Letters}, author={Li, Chi-Kang and Petrasso, Richard D.}, year={1993}, month=may, pages={3063–3066}, language={en} }

@article{Zylstra_Hurricane_2019, title={On alpha-particle transport in inertial fusion}, volume={26}, ISSN={1070-664X}, DOI={10.1063/1.5101074}, abstractNote={Analytical theory and models of inertial fusion implosion use parameterized microphysics models. In this paper, we consider the DT alpha-particle transport, and report new parameterizations of the range, heating efficacy, and energy partition using modern stopping-power theory. Our resulting heating efficacy is lower than previously published results, which reduces the temperature and pressure generated by a dynamic implosion hot-spot evolution model, and shifts the burning-plasma regime boundary slightly farther from current experimental results.}, number={6}, journal={Physics of Plasmas}, author={Zylstra, A. B. and Hurricane, O. A.}, year={2019}, month=june, pages={062701} }

@article{Son_Fisch_2005, title={Current-Drive Efficiency in a Degenerate Plasma}, volume={95}, DOI={10.1103/PhysRevLett.95.225002}, abstractNote={In a degenerate plasma, the rates of electron processes are much smaller than the classical model would predict, affecting the efficiencies of current generation by external noninductive means, such as by electromagnetic radiation or intense ion beams. For electron-based mechanisms, the current-drive efficiency is higher than the classical prediction by more than a factor of 6 in a degenerate hydrogen plasma, mainly because the electron-electron collisions do not quickly slow down fast electrons. Moreover, electrons much faster than thermal speeds are more readily excited without exciting thermal electrons. In ion-based mechanisms of current drive, the efficiency is likewise enhanced due to the degeneracy effects, since the electron stopping power on slow ion beams is significantly reduced.}, number={22}, journal={Physical Review Letters}, publisher={American Physical Society}, author={Son, S. and Fisch, N. J.}, year={2005}, month=nov, pages={225002} }

@article{Malko_et_2022, title={Proton stopping measurements at low velocity in warm dense carbon}, volume={13}, rights={2022 This is a U.S. government work and not under copyright protection in the U.S.; foreign copyright protection may apply}, ISSN={2041-1723}, DOI={10.1038/s41467-022-30472-8}, abstractNote={Ion stopping in warm dense matter is a process of fundamental importance for the understanding of the properties of dense plasmas, the realization and the interpretation of experiments involving ion-beam-heated warm dense matter samples, and for inertial confinement fusion research. The theoretical description of the ion stopping power in warm dense matter is difficult notably due to electron coupling and degeneracy, and measurements are still largely missing. In particular, the low-velocity stopping range, that features the largest modelling uncertainties, remains virtually unexplored. Here, we report proton energy-loss measurements in warm dense plasma at unprecedented low projectile velocities. Our energy-loss data, combined with a precise target characterization based on plasma-emission measurements using two independent spectroscopy diagnostics, demonstrate a significant deviation of the stopping power from classical models in this regime. In particular, we show that our results are in closest agreement with recent first-principles simulations based on time-dependent density functional theory.}, number={11}, journal={Nature Communications}, publisher={Nature Publishing Group}, author={Malko, S. and Cayzac, W. and Ospina-Bohórquez, V. and Bhutwala, K. and Bailly-Grandvaux, M. and McGuffey, C. and Fedosejevs, R. and Vaisseau, X. and Tauschwitz, An and Apiñaniz, J. I. and others}, year={2022}, month=may, pages={2893}, language={en} }

@article{VanZeeland_et_2021, title={Beam modulation and bump-on-tail effects on Alfvén eigenmode stability in DIII-D}, volume={61}, ISSN={0029-5515}, DOI={10.1088/1741-4326/abf174}, abstractNote={Beam modulation effects on Alfvén eigenmode stability have been investigated in a recent DIII-D experiment and show that variations in neutral beam modulation period can have an impact on the beam driven Alfvén eigenmode spectrum and resultant fast ion transport despite similar time-averaged input power. The experiment was carried out during the current ramp phase of L-mode discharges heated with sub-Alfvénic 50–80 kV deuterium neutral beams that drive a variety of Alfvén eigenmodes unstable. The modulation period of two interleaved beams with different tangency radii was varied from shot to shot in order to modify the relative time dependent mix of the beam pitch angle distribution as well as the persistence of a bump-on-tail feature near the injection energy (a feature confirmed by imaging neutral particle analyzer measurements). As the beam modulation period is varied from 7 ms to 30 ms on/off (typical full energy slowing down time of τ slow ≈ 50 ms at mid-radius), toroidicity-induced Alfvén eigenmodes (TAEs) located in the outer periphery of the plasma become intermittent and coincident with the more tangential beam. Core mode activity changes from reversed shear Alfvén eigenmodes (RSAEs) to a mix of RSAE and beta-induced Alfvén eigenmodes. Discharges with 30 ms on/off period do not have a persistent bump-on-tail feature, have the lowest average mode amplitude and least fast ion transport. Detailed analysis of an individual TAE using TRANSP kick modeling (Monte Carlo evolution of the distribution function with probabilistic ‘kicks’ by the AEs) and the resistive MHD code with kinetic fast ions, MEGA, find no strong role of energy gradient drive due to bump-on-tail features. Instead, the observed TAE modulation with interleaved beams is likely a pitch angle dependent result combined with slowing down of the tangential beam between pulses. For the conditions investigated, bump-on-tail contributions to TAE drive were found to be 5% or less of the total drive at any given time.}, number={6}, journal={Nuclear Fusion}, publisher={IOP Publishing}, author={Van Zeeland, M.A. and Bardoczi, L. and Gonzalez-Martin, J. and Heidbrink, W.W. and Podesta, M. and Austin, M. and Collins, C.S. and Du, X.D. and Duarte, V.N. and Garcia-Munoz, M. and others}, year={2021}, month=may, pages={066028}, language={en} }

@article{Graves_Chapman_Coda_Lennholm_Albergante_Jucker_2012, title={Control of magnetohydrodynamic stability by phase space engineering of energetic ions in tokamak plasmas}, volume={3}, rights={2012 Springer Nature Limited}, ISSN={2041-1723}, DOI={10.1038/ncomms1622}, abstractNote={Virtually collisionless magnetic mirror-trapped energetic ion populations often partially stabilize internally driven magnetohydrodynamic disturbances in the magnetosphere and in toroidal laboratory plasma devices such as the tokamak. This results in less frequent but dangerously enlarged plasma reorganization. Unique to the toroidal magnetic configuration are confined “circulating” energetic particles that are not mirror trapped. Here we show that a newly discovered effect from hybrid kinetic-magnetohydrodynamic theory has been exploited in sophisticated phase space engineering techniques for controlling stability in the tokamak. These theoretical predictions have been confirmed, and the technique successfully applied in the Joint European Torus. Manipulation of auxiliary ion heating systems can create an asymmetry in the distribution of energetic circulating ions in the velocity orientated along magnetic field lines. We show the first experiments in which large sawtooth collapses have been controlled by this technique, and neoclassical tearing modes avoided, in high-performance reactor-relevant plasmas.}, number={1}, journal={Nature Communications}, publisher={Nature Publishing Group}, author={Graves, J. P. and Chapman, I. T. and Coda, S. and Lennholm, M. and Albergante, M. and Jucker, M.}, year={2012}, month=jan, pages={624}, language={en} }

@article{Qin_2024, title={Advanced fuel fusion, phase space engineering, and structure-preserving geometric algorithms}, volume={31}, ISSN={1070-664X}, DOI={10.1063/5.0203707}, abstractNote={Non-thermal advanced fuel fusion trades the requirement of a large amount of recirculating tritium in the system for that of large recirculating power. Phase space engineering technologies utilizing externally injected electromagnetic fields can be applied to meet the challenge of maintaining non-thermal particle distributions at a reasonable cost. The physical processes of the phase space engineering are studied from a theoretical and algorithmic perspective. It is emphasized that the operational space of phase space engineering is limited by the underpinning symplectic dynamics of charged particles. The phase space incompressibility according to the Liouville theorem is just one of many constraints, and Gromov’s non-squeezing theorem determines the minimum footprint of the charged particles on every conjugate phase space plane. In this sense and level of sophistication, the mathematical abstraction of phase space engineering is symplectic topology. To simulate the processes of phase space engineering, such as the Maxwell demon and electromagnetic energy extraction, and to accurately calculate the minimum footprints of charged particles, recently developed structure-preserving geometric algorithms can be used. The family of algorithms conserves exactly, on discretized spacetime, symplecticity and thus incompressibility, non-squeezability, and symplectic capacities. The algorithms apply to the dynamics of charged particles under the influence of external electromagnetic fields as well as the charged particle–electromagnetic field system governed by the Vlasov–Maxwell equations.}, number={5}, journal={Physics of Plasmas}, author={Qin, Hong}, year={2024}, month=may, pages={050601} }

@article{Updike_Bohlsen_Qin_Fisch_2025, title={Minimizing phase-space energies}, volume={112}, DOI={10.1103/4rbz-cgxf}, abstractNote={A primary technical challenge for harnessing fusion energy is to control and extract energy from a nonthermal distribution of charged particles. The fact that phase space evolves by symplectomorphisms fundamentally limits how a distribution may be manipulated. While the constraint of phase-space volume preservation is well understood, other constraints remain to be fully appreciated. To better understand these constraints, we study the problem of extracting energy from a distribution of particles using area-preserving and symplectic linear maps. When a quadratic potential is imposed, we find that the maximal extractable energy can be computed as trace minimization problems. We solve these problems and show that the extractable energy under linear symplectomorphisms may be much smaller than the extractable energy under special linear maps. The method introduced in the present study enables an energy-based proof of the linear Gromov nonsqueezing theorem.}, number={3}, journal={Physical Review E}, publisher={American Physical Society}, author={Updike, Michael and Bohlsen, Nicholas and Qin, Hong and Fisch, Nathaniel J.}, year={2025}, month=sept, pages={035202} }

@article{Fisch_1987, title={Theory of current drive in plasmas}, volume={59}, DOI={10.1103/RevModPhys.59.175}, abstractNote={The continuous operation of a tokamak fusion reactor requires, among other things, a means of providing continuously the toroidal current. Such operation is preferred to the conventional pulsed operation, where the plasma current is induced by a time-varying magnetic field. A variety of methods have been proposed to provide continuous current, including methods that utilize particle beams or radio-frequency waves in any of several frequency regimes. Currents as large as half a mega-amp have now been produced in the laboratory by such means, and experimentation in these techniques has now involved major tokamak facilities worldwide.}, number={1}, journal={Reviews of Modern Physics}, publisher={American Physical Society}, author={Fisch, Nathaniel J.}, year={1987}, month=jan, pages={175–234} }

@article{Kolmes_Fisch_2024, title={Upper and lower bounds on phase-space rearrangements}, volume={31}, ISSN={1070-664X}, DOI={10.1063/5.0202456}, abstractNote={Broad classes of plasma phenomena can be understood in terms of phase-space rearrangements. For example, the net effect of a wave–particle interaction may consist of moving populations of particles from one region of phase space to another. Different phenomena drive rearrangements that obey different rules. When those rules can be specified, it is possible to calculate bounds that limit the possible effects the rearrangement could have (such as limits on how much energy can be extracted from the particles). This leads to two problems. The first is to understand the mapping between the allowed class of rearrangements and the possible outcomes that these rearrangements can have on the overall distribution. The second is to understand which rules are appropriate for which physical systems. There has been recent progress on both fronts, but a variety of interesting questions remain unanswered.}, number={4}, journal={Physics of Plasmas}, author={Kolmes, E. J. and Fisch, N. J.}, year={2024}, month=apr, pages={042109} }

@article{Qin_Kolmes_Updike_Bohlsen_Fisch_2025, title={Gromov ground state in phase space engineering for fusion energy}, volume={111}, DOI={10.1103/PhysRevE.111.025205}, abstractNote={Phase space engineering by rf waves plays important roles in both thermal D-T fusion and nonthermal advanced fuel fusion, but not all phase space manipulation is allowed; certain fundamental limits exist. In addition to Liouville’s theorem, which requires the manipulation to be volume preserving, Gromov’s nonsqueezing theorem imposes another constraint. The Gardner ground state is defined as the ground state accessible by smooth volume-preserving maps. However, the extra Gromov constraint should produce a higher-energy ground state. An example of a Gardner ground state forbidden by Gromov’s nonsqueezing theorem is given. The challenge question is “What is the Gromov ground state, i.e., the lowest energy state accessible by smooth symplectic maps?” This is a difficult problem. As a simplification, we conjecture that the linear Gromov ground state problem is solvable.}, number={2}, journal={Physical Review E}, publisher={American Physical Society}, author={Qin, Hong and Kolmes, Elijah J. and Updike, Michael and Bohlsen, Nicholas and Fisch, Nathaniel J.}, year={2025}, month=feb, pages={025205} }
